\documentclass[journal]{new-aiaa}
\usepackage[utf8]{inputenc}
\usepackage{xcolor}
\usepackage{graphicx}
\usepackage{amsmath,doi}
\usepackage{verbatim}
\usepackage[version=4]{mhchem}
\usepackage{siunitx}
\usepackage{longtable,tabularx}
\usepackage{subcaption}
\DeclareCaptionSubType*{figure}

\newcommand\norm[1]{\left\lVert#1\right\rVert}
\newcommand\abs[1]{\left|#1\right|}
\DeclareMathOperator*{\argmin}{arg\,min} 
\title{Feasibility analysis of ensemble sensitivity computation in turbulent flows}

\author{Nisha Chandramoorthy \footnote{Ph.D. Candidate, Department of Mechanical Engineering, nishac@mit.edu, AIAA Student Member} and Pablo Fernandez\footnote{Ph.D. Candidate, Department of Aeronautics and Astronautics, pablof@mit.edu, AIAA Student Member}}
\affil{Massachusetts Institute of Technology, Cambridge, MA, 02139, USA}
\author{Chaitanya Talnikar\footnote{Developer technology engineer}}
\affil{NVIDIA Corporation, Santa Clara, CA, 95051, USA}
\author{Qiqi Wang\footnote{Associate Professor, Department of Aeronautics and Astronautics, qiqi@mit.edu, AIAA Associate Fellow}}
\affil{Massachusetts Institute of Technology, Cambridge, MA, 02139, USA}

\begin{document}

\maketitle

\begin{abstract}
In chaotic systems, such as turbulent flows, the solutions to tangent
and adjoint equations exhibit an unbounded growth in their norms.
This behavior renders the instantaneous tangent and adjoint solutions 
unusable for sensitivity analysis. The Lea-Allen-Haine ensemble sensitivity (ES) estimates provide a way of computing meaningful sensitivities
in chaotic systems by utilizing tangent/adjoint solutions 
over short trajectories. In this paper, 
we analyze the feasibility of ES computations under optimistic mathematical assumptions
on the flow dynamics. Furthermore, we estimate upper bounds on the rate of convergence 
of the ES method in numerical simulations of turbulent flow. Even at the optimistic 
upper bound, the ES method is computationally intractable in each of the numerical examples considered.
\end{abstract}

\section*{Nomenclature}

{\renewcommand\arraystretch{1.0}
\noindent\begin{longtable*}{@{}l @{\quad \quad} l@{}}
ES & Ensemble Sensitivity \\
$\tau$  & trajectory length for ES computation \\
$N$ &    number of i.i.d samples used in ES computation \\
$\theta_{\tau,N}$& ES estimator \\
	&Superscripts 
	T, A and FD stand for tangent, 
	adjoint and finite difference respectively\\
$u$ & a $d$-dimensional state (or phase) vector   \\
	& used to represent a primal state \\
$s$ & scalar parameter \\
	$u(t,s,u_0)$ & primal state vector at time $t$ with initial state $u_0$ \\
$u(t,s),\: u(t)$ & this notation is used to represent the primal state when the \\
	&initial state and/or the parameter need not be explicitly mentioned\\
$J(u(t,s,u_0))$ & objective function evaluated at state $u(t)$ \\
$y(t, u_0)$ & adjoint solution at time $t$ corresponding to \\
	&primal solution starting at initial condition $u_0$ \\
$v(t, u_0)$ & tangent solution at time $t$ corresponding to  \\
	&primal solution starting at initial condition $u_0$ \\
$v^h(t,u_0)$ & homogeneous tangent solution corresponding to \\
	& primal initial condition $u_0$ \\
$f (u(t),s)$ & the vector field on the right hand side of the primal set of ODEs, \\
	&evaluated at the state $u(t)$\\
$\dfrac{\partial g}{\partial u}\Big|_t$ & 
	partial derivative of a scalar or a vector field $g(u(t),s)$ \\
	&with respect to $u(t)$, evaluated at $(u(t),s),$\\
	& where $s$ is the reference value of the parameter; \\
	&e.g., $(\partial f/\partial u)\Big|_0$ indicates the Jacobian matrix at $u_0$.\\
$\dfrac{\partial g}{\partial u_0}\Big|_t$ & 
	partial derivative of a scalar or a vector field $g(u(t,s,u_0),s)$ \\
	&with respect to $u_0$, evaluated at $(u(t,s,u_0),s),$\\
	& where $s$ is the reference value of the parameter \\
$\dfrac{\partial g}{\partial s}\Big|_t$ & partial derivative with respect to $s$, 
	of a scalar or a vector field $g$,\\
	&evaluated at $(u(t),s)$, where $s$ is \\
	&the reference value of the parameter \\
$\mu_s$ & stationary measure the statistics wrt which we are interested in \\
	$\mu_s( J)$ & phase-space average, according to $\mu_s$, of the function $J$ \\
	${\rm b}(\tau)$ & bias in $\theta_{\tau,N}$ \\
	${\rm var}(\tau,N)$ or ${\rm var}(\theta_{\tau,N})$ & variance of $\theta_{\tau,N}$ \\
$\lambda_1$ & largest Lyapunov exponent \\
	$\gamma_1$ & rate of exponential decay of ${\rm b}(\tau)$\\
	$f(x) \sim {\cal O}(g(x))$ & indicates that a function $f$ is  \\
	&on the same order as $g$. That is, there exists  
	an $x^*\geq 0$ such that \\
	&$\abs{f(x)} \leq C \abs{g(x)}$, 
	for all $\abs{x} \geq x^*$, 
	with the constant $C > 0$ \\
	&being independent of $x$. 
\end{longtable*}}

\section{Introduction}
\label{ref:introduction}
\lettrine{G}{radient}-based computational approaches in multi-disciplinary design and optimization require
sensitivity information computed from numerical simulations of fluid flow. In Reynolds-averaged-Navier-Stokes (RANS)
simulations, sensitivity computation is traditionally performed using tangent or adjoint equations or using
finite difference methods. Sensitivities computed from RANS simulations and 
non-chaotic Navier-Stokes simulations have been extensively applied toward 
uncertainty quantification \citep{palacios, qiqi-unstart}, mesh adaptation \citep{fidkowski}, flow control
\cite{rizzetta}, noise reduction \citep{bodony, engblom, paul}  and aerostructural design
optimization applications \citep{peter,giles1,
giles2, alonso2, nielsen}. 
Many modern applications require computing sensitivities in direct numerical simulations (DNS) or 
large-eddy simulations (LES); examples include buffet prediction in high-maneuverability 
aircraft, modern turbomachinery design and jet engine and
airframe noise control. Conventional tangent/adjoint approaches cannot be used to compute sensitivities
of statistical averages (or long-time averaged quantities) in these high-fidelity, eddy-resolving simulations. This is because the tangent and adjoint solutions diverge exponentially \citep{angxiu,lea}, since these
simulations exhibit chaotic behavior i.e., infinitesimal perturbations to initial conditions grow exponentially in time. For this reason, sensitivity studies on DNS or LES have been
restricted to short-time horizons; these short-time sensitivities have found limited applicability 
including in flow control in combustion systems \citep{capecelatro}, jet noise reduction \citep{bodony}
and structural design \citep{alonso1}.  

The first proposed approach to tackle the computation of sensitivities
of statistics is the Lea-Allen-Haine ensemble sensitivity (ES) method \citep{lea}. 
In this method, the problem of exponentially diverging linearized 
perturbation solutions (such as tangent/adjoint) is mitigated by taking a sample
average of short-time sensitivities. The method approximates Ruelle's 
response formula \citep{ruelle} for sensitivity of average quantities to system parameters. 
The convergence of the method has been 
shown by Eyink \emph{et al.} \cite{eyink} in the limit of taking an infinite number of samples and 
increasing the time duration for sensitivity computation to infinity, in that order. Eyink \emph{et al.} \cite{eyink} 
establish that the rate of convergence is worse than a typical Monte Carlo simulation 
(in which the error in a sample average diminishes at the rate $1/\sqrt{N}$, where $N$ is the number of samples) in the case
of the classical 3-variable Lorenz'63 system. However, the convergence trend is still unknown for general
chaotic systems. In this work, we present an analysis of the mean squared error of the ES method 
as a function of the computational cost for a certain class of systems called uniformly hyperbolic systems \cite{katok}. 
It is worth noting that at the time of writing of this paper, alternatives to the ES method \cite{angxiu_nilsas,lucarini} are under active investigation.
Non-intrusive least squares shadowing (NILSS) \cite{angxiu, patrick} and its adjoint-variant \citep{angxiu_nilsas, patrick} are methods
that are conceptually based on the shadowing property of uniformly hyperbolic systems. This property enables the 
computation of a particular tangent solution that remains bounded in a long time window and sensitivities are estimated
using this tangent solution. The NILSS algorithm requires the knowledge of the unstable subspace (of the tangent 
space, corresponding to the positive Lyapunov exponents). This makes
the algorithm expensive when the dimension of the subspace
or the number of positive Lyapunov exponents is large. The NILSS method has been applied to LES of turbulent flows around bluff bodies \citep{patrickPablo, patrick-aiaa} at low Reynolds 
numbers, where the number of positive Lyapunov exponents is small enough to limit the computational expense when compared to wall-bounded
flows, for instance.

The paper is organized as follows. In the next section, we review the ES method and define the ES estimator for the sensitivity. In section 
\ref{sec:analysis}, we describe the mean squared error in the ES estimator in terms of the associated bias and variance 
and obtain optimistic, problem-dependent estimates for both components. We predict the rate of convergence
as a function of computational cost under these optimistic assumptions on the dynamics. The rest of the paper
consists of numerical examples that illustrate the convergence trend of the ES method. In sections \ref{sec:lorenz63} and
\ref{sec:lorenz96}, we discuss two low-dimensional models of chaotic fluid behavior: the Lorenz'63  and 
Lorenz'96 systems. We apply our optimistic analysis to roughly estimate an upper bound on the 
rate of convergence. We then present two numerical simulation results that serve to illustrate the applicability of
ES schemes in fluid simulations of practical interest, in light of our mathematical analysis in section \ref{sec:analysis}. 
The first is a simulation of a NACA 0012 airfoil in section \ref{sec:airfoil} and the second, an LES of turbulent
flow around a turbine vane in section \ref{sec:vane}. Section \ref{sec:discussion} contains some practical 
recommendations on the applicability 
of the ES method, based on analytical and numerical insights from the previous sections. 

\section{The ES estimator}
\label{sec:algorithm}
\subsection{The sensitivity computation problem setup}
Consider a chaotic fluid flow parameterized by $s$, expressed through a PDE system
of the following form:
\begin{align}
	\label{eqn:primal}
	\frac{\partial u(t,s)}{\partial t} &= f(u(t,s),s) \;\;\qquad u(t,s)\in \mathbb{R}^d, 
	{\color{black}s\in \mathbb{R}^p}, \\
	\notag
	u(0,s) &= u_0 \in \mathbb{R}^d.		
\end{align}
Here, we use $u(t,s)$ to denote the state vector at time $t$ 
obtained due to the evolution of an initial state $u_0$ according to Eq. 
\ref{eqn:primal}; $u_0$ is chosen independently of $s$.
As an example, if Eq.\ref{eqn:primal} is the spatially discretized
incompressible Navier-Stokes equation, a state vector consists of the velocity components, 
and pressure at all the grid points. In this case, the dimension
of the parameterized, spatially-discretized Navier-Stokes 
state vector -- $u(t,s)$ -- is $d = 4\times$ the number of degrees of
freedom (and in a two-dimensional flow, $d=3\times $ the degrees of freedom). The state vector $u(t,s)$
is approximately known from numerical simulation and can be thought of as 
a point in the phase space $M$, a compact subset of $\mathbb{R}^d$. The set of input parameters $s$ can 
be, for instance, related to the inlet conditions, the geometry of the domain or solid 
bodies in the flow and so on. In the interest of simplicity, from here on, $s$ is a scalar parameter.
The state vector is also a function of the initial condition and to make 
explicit this dependence, we write as $u(t, s, u_0)$ the solution of 
Eq.\ref{eqn:primal} at time $t$, solved with $u_0$ as the initial state. 

The fluid flows we consider here are statistically stationary, i.e., the states
in phase space are distributed according to a time-invariant probability distribution $\mu_s$. The subscript $s$ indicates that the stationary distribution is 
a function of the parameter; we work under the assumption that $\mu_s$ 
is a smooth function of $s$. Suppose $J$ is a smooth scalar function of the state such as the lift/drag ratio or the 
pressure loss in a turbine wake. The ensemble mean of $J$ is defined as its expectation with respect to 
$\mu_s$ and denoted by $\mu_s( J) := \int J \; d\mu_s$. We are interested in computing
the sensitivity of $\mu_s( J)$ to $s$. The ensemble mean $\mu_s(J)$, is an average over phase space,
and hence can be measured as a time average 
along almost every trajectory, under the assumption of ergodicity. 
More precisely, for almost every initial state $u_0$,
\begin{align*}
	\mu_s( J) = \lim_{t\to\infty} \dfrac{1}{t} 
	\int_0^t J(u(t',s, u_0))\; dt'. 
\end{align*}
In practice, $\mu_s( J )$ is computed approximately
as a finite-time average by truncating a trajectory
at a large $t$. The Lea-Allen-Haine ES method, the subject of this paper,
computes the sensitivity $(d\mu_s( J)/ds)$ approximately, as we describe next.
\subsection{The Lea-Allen-Haine ES algorithm }
\label{sec:algorithm}
The ES estimator of the sensitivity
is the sample mean of a finite number of
independent sensitivity outputs
computed over short trajectories. We now make this statement 
precise in the following description of the ES algorithm.
Consider $N$ independent
initial states $\left\{ u_0^{(i)}\right\}_{i=1}^N$, sampled according to $\mu_s$. 
Let us denote the sensitivity computed along a 
flow trajectory, 
of length $\tau$, starting from $u_0^{(i)}$, as $\theta_\tau^{(i)}$. Then,
the Lea-Allen-Haine estimator $\theta_{\tau,N}$, is given by,
\begin{align}
	\theta_{\tau,N} &= \frac{1}{N} \sum_{i=1}^{N} \theta_\tau^{(i)}. 	\label{eqn:TimeAveragedSensitivity}
\end{align}
 
The standard adjoint method was proposed to be used originally \cite{lea,eyink} to 
compute the sensitivities
$\theta_\tau^{(i)}$ in order to retain the advantage of adjoint methods, namely
that they scale well with the parameter space dimension, since the parameters 
enter the computation only when determining the sensitivities using the 
adjoint solution vectors, and the adjoint vectors the computation of which 
is the majority of the computational cost, are themselves parameter-independent. The analysis 
in the remainder of this paper would be
identical
however, if the tangent equation or a finite difference approximation
to the sensitivity derivative was used instead. 
The three methods of computing $\theta_\tau^{(i)}$, 
dropping the superscript $i$ for clarity,
are listed below. The sensitivity estimator computed using Eq.
\ref{eqn:TimeAveragedSensitivity} when $\theta_\tau^{(i)}$ are computed
using the tangent equation, adjoint equation and from finite difference
are denoted using $\theta_{\tau,N}^{\rm T}$,  $\theta_{\tau,N}^{\rm A}$
and  $\theta_{\tau,N}^{\rm FD}$ respectively. 
\begin{enumerate}
\item From the tangent equation:
		\begin{align}
			\frac{d v(t, u_0)}{dt} = 
			\frac{\partial f}{\partial s}\Big|_t + 
			\frac{\partial f}{\partial u}\Big|_t\; v(t, u_0)\;\;		
				\label{eqn:TangentSystem}
		\end{align}
		where, $v(t, u_0) = (du(t,s,u_0)/ds)$ is the tangent solution 
        at time $t$ when the primal initial condition is $u_0$. The tangent 
		initial condition $v(0, u_0) = 0 \in \mathbb{R}^d$ 
		since $u_0$ is independent of $s$. We use $({\partial f}/{\partial s})\Big|_t$ 
		to represent the partial derivative of $f$ with respect to the
		parameter $s$, evaluated at $u(t,s,u_0)$. The Jacobian matrix
		at time $t$ is written as $(\partial f/\partial u)\Big|_t$. The 
		ES estimator based on the tangent equation is
		\begin{align}
			\theta_{\tau,N}^{\rm T} 
			&=\frac{1}{\tau N} 
			\sum_{i=1}^N 
				\int_0^\tau  
				\frac{\partial J}{\partial u}\Big|_{t,i}
			\cdot  v(t, u_0^{(i)}) \; dt,
				\label{eqn:TangentSystem2}
		\end{align}
		where we have used a second subscript $i$ in $\dfrac{\partial J}{\partial u}\Big|_{t,i}$
		to indicate that the primal initial condition is $u_0^{(i)}$.
		\item From the adjoint equation: 
		\begin{align}
			\frac{d y(t,u_0)}{d t} &= -\frac{\partial J}{\partial u}\Big|_t
				\; -\; \left(\frac{\partial f}{\partial u}\right)^*\Big|_t
				\;y(t,u_0)
		 ,
		\label{eqn:AdjointSystem}
		\end{align}
		where $y(\tau, u_0) = 0 \in \mathbb{R}^d$ is the adjoint vector at time $\tau$ 
		which results in the adjoint vector $y(t,u_0)$ at time $t$ 
		when evolved backwards in time for a time $\tau - t$.  
		Here $(\partial f/\partial u)^*\Big|_t$
		is the conjugate transpose of the Jacobian 
		matrix at time $t$. The ES estimator 
		based on the adjoint equation is,
	\begin{align}
		\theta^{\rm A}_{\tau,N} &=
		\frac{1}{\tau N} \sum_{i=1}^N
		\int_0^\tau 
		y(t, u_0^{(i)})  
		\cdot \frac{\partial f}{\partial s}\Big|_{t,i}\; dt,
		\label{eqn:AdjointSystem2}
		\end{align}
	where we have used a second subscript $i$ in $(\partial f/\partial s)\Big|_{t,i}$ to indicate that the primal initial condition is 
		$u_0^{(i)}$.
	\item Using, for example, a second-order accurate finite difference approximation:
		\begin{align}
			\theta^{\rm FD}_{\tau, N}= \frac{1}{\tau N} 
			\sum_{i=1}^N\frac{1}{2 \Delta s} \Bigg(
			\int_0^\tau J(u(t,s + \Delta s,u_0^{(i)})) \; dt  - 
			\int_0^\tau J(u(t,s - \Delta s,u_0^{(i)}))	\; dt\Bigg),
			\label{eqn:FiniteDifference}
		\end{align}
		for a small $\Delta s$.
\end{enumerate}

\section{Error vs. computational cost of the ES estimator}

Chaotic systems such as turbulent fluid flows often 
exhibit regularity in long-time averages despite showing  
seeming randomness in instantaneous measurements. The chaotic 
hypothesis, proposed by Gallavotti and Cohen \cite{gallavotti}, is 
the notion that these systems can be treated, for the purpose of 
studying their long-time behavior, as having a certain smooth structure in phase space. 
This smooth structure allows for the existence of subspaces of the 
tangent space consisting of expanding and contracting derivatives of state functions. 
Our goal in this section is to predict the convergence trend 
of the ES method for systems that satisfy the chaotic hypothesis.  
More specifically, we would like to construct 
an optimistic model for the least mean squared error of the 
ES method, achievable at a given 
computational cost, in these systems. Before we delve into the construction of 
the optimistic model, we will discuss the rigorous justification
for the convergence of the ES method -- Ruelle's response formula. 
Here we will focus on the implications of the formula 
for the ES method without going into details; 
the reader is referred
to Ruelle's original paper \cite{ruelle,ruelle1} for the derivation of the 
response formula. We shall refrain from a technical discussion on 
hyperbolicity and other concepts from dynamical systems theory but provide
the necessary qualitative description, in the context used here.

\label{sec:analysis}
\subsection{ES estimator as approximation of Ruelle's response formula}
 \label{subsec:SufficientConditions}

The following response formula \cite{ruelle_hyperbolicflows} due to 
Ruelle gives the sensitivity to $s$, of an objective function $J$'s
statistical average,
\begin{align}
	\frac{d \mu_s( J)}{ds} = \int_0^\infty dt \int_M 
	\frac{\partial J}{\partial u_0}\Big|_t \cdot 
	\frac{\partial f}{\partial s}\Big|_0
	\; d\mu_s (u_0),
	\label{eqn:ruelleLinearResponse}
\end{align}
where $(\partial J/\partial u_0)\Big|_t := 
\partial J(u(t,s;u_0))/\partial u_0$ represents the derivative
of $J$ at time $t$ with respect to 
the initial state $u_0$. One can interpret the 
inner integral (over $M$) as the 
statistical response
of the objective function at time instant $t$. It is a phase space 
average of the sensitivity at time $t$ with the initial conditions 
distributed according to the 
stationary probability distribution $\mu_s$. 
Ruelle's formula has been proven to hold for a certain class 
of smooth dynamical systems known as uniformly hyperbolic systems.
Roughly speaking, these are systems in which the tangent space at 
every point in phase space can be split into stable and unstable 
subspaces, which contain respectively, exponentially decaying and 
growing tangent vectors. In reality, the formula is 
applicable to a large class of fluid flow problems (this wider applicability referenced earlier
as the chaotic hypothesis) that are not necessarily uniformly hyperbolic, 
as subsequent works \cite{gallavotti,ruelle}
have analyzed.

An iterated integral such as in Eq. \ref{eqn:ruelleLinearResponse},
gives the same value upon 
switching the order of integration if and only if the double integral, in which the integrand
is replaced by its absolute value, is finite (this is the Fubini-Tonelli theorem). 
The absolute value of the integrand in Eq. \ref{eqn:ruelleLinearResponse}  
diverges to infinity as $t \to \infty$ for almost every initial condition in phase space.
In other words, in a chaotic system, 
for every initial condition $u_0$ in phase space except those in a set of 
$\mu_s$-measure 0, 
$$ \int_0^\infty \abs{\frac{\partial J}{\partial u_0}\Big|_t \cdot \frac{\partial f}{\partial s}\Big|_0
} \: dt 
= \infty.$$

For this reason, in Eq.\ref{eqn:ruelleLinearResponse}, the integral over phase space and over 
time do not commute. The iterated integral in Eq.\ref{eqn:ruelleLinearResponse}
in which the integration over phase space
is performed first, leads to a bounded value, which is equal to $d\mu_s( J)/ds$. 
On the other hand, changing the order
of integration and integrating over $t$ first, results in infinity. 

From here on, we use the following sample average as the definition of the ES estimator
\begin{align}
	\theta_{\tau,N} := \frac{1}{N} \sum_{i=1}^{N}
	\int_0^\tau \frac{\partial J}{\partial u_0}\Big|_{t,i}\;\cdot\; \frac{\partial f}{\partial s}\Big|_{0,i}\;\; dt,
	\label{eqn:estimatorFormula}
\end{align}
where the set of initial states $\left\{ u^{(i)}_0 \right\}_{i=1}^N$ {\color{blue} is} independent
and identically distributed according to $\mu_s$. The ES
estimator can be interpreted as an
approximation of Ruelle's formula in the following sense: if the outer integral
over time (in Eq. \ref{eqn:ruelleLinearResponse}) is truncated at time $\tau$
and the phase space average of the integrand approximated with a sample mean over $N$ independent
samples, we obtain the estimator in Eq. \ref{eqn:estimatorFormula}. The definition can 
also be interpreted as the average sensitivity of $\int_0^\tau  J(u(t,s,u_0)) \; dt$ to 
initial condition perturbations along the vector field
$(\partial f/\partial s)\Big|_0$. This can be computed
using the tangent equation method listed in \ref{sec:algorithm}
but using the homogeneous tangent equation -- i.e., the tangent
equation without the forcing term, $(\partial f/\partial s)\Big|_t$. 
To wit, the homogeneous 
tangent equation solved with the initial condition 
$v^h(0, u_0) = (\partial f/\partial s)\Big|_0$ yields at time $t$, the solution,
$$ v^h(t, u_0) = \frac{\partial u}{\partial u_0}\Big|_t\;\dfrac{\partial f}{\partial s}\Big|_0.$$
Therefore, the sensitivity of $\int_0^\tau J(u(t,s,u_0))\; dt$ to 
initial condition perturbations along $(\partial f/\partial s)\Big|_0$
is,
\begin{align*}
	\int_0^\tau \dfrac{\partial J}{\partial u_0}\Big|_t \cdot
	\dfrac{\partial f}{\partial s}\Big|_0 \; dt &= 
	\int_0^\tau \frac{\partial J}{\partial u}\Big|_t
	\: \cdot \: \frac{\partial u}{\partial u_0}\Big|_t 
	\; \frac{\partial f}{\partial s}\; dt =  
	\int_0^\tau \frac{\partial J}{\partial u}\Big|_t 
	\; \cdot \; v^h(t,u_0). 
\end{align*}
Taking an $N$-sample average of the above results in the formula for the estimator \ref{eqn:estimatorFormula}.
The resulting $\theta_{\tau, N}$ is different from the sensitivities $\theta_{\tau,N}^{\rm FD}$,
$\theta_{\tau,N}^{\rm T}$ and $\theta_{\tau,N}^{\rm A}$ defined in section \ref{sec:algorithm}, all
of which are also different from one another. However, in the asymptotic limit of $\tau \to \infty$, all these sensitivities
grow exponentially ($\theta_{\tau,N}^{\rm FD}$ does not grow unbounded in norm with $\tau$, but rather saturates, for 
a non-zero value of $\Delta s$) at the same rate determined by the largest
among the Lyapunov exponents (the asymptotic 
exponential growth or decay rate of tangent/adjoint vectors) of the system. Therefore, for the purpose of an asymptotic analysis, 
we restrict our attention to the estimator defined by
Eq. \ref{eqn:estimatorFormula} and refer to $\theta_{\tau,N}$ using the umbrella term ES estimator.

One notices that in the practical computation of the ES method, the integrals are commuted 
when compared to Ruelle's formula but the integral in time is truncated at a finite time. 
As noted in Eyink et al's analysis \cite{eyink}, the rationale behind 
swapping the order of integration as compared to Ruelle's formula in Eq.
\ref{eqn:ruelleLinearResponse} is the observation that the ``divergence of the individual adjoints is delayed
on taking a sample average'', for the Lorenz'63 system, a low-order model for fluid convection that we discuss
in section \ref{sec:lorenz63}. In the rest of the paper, our goal is to analyze
the convergence trend in more generality.

\subsection{Bias and variance of the ES estimator}
\label{sec:optimisticEstimates}
Having defined the ES estimator, we now construct an optimistic 
model for its mean squared error in uniformly hyperbolic systems. 
In this section, we present our choices of 
optimistic estimates for the bias and 
the variance associated with the estimator and use uniform hyperbolicity to 
justify those choices. The ES estimator, as defined in Eq.\ref{eqn:estimatorFormula},
has a non-zero bias for a finite $\tau$. By definition, the bias, 
denoted by ${\rm b}(\tau)$, gives the difference between the value attained 
by the estimator on using an infinite number of samples and the true value
of the sensitivity,
\begin{align}
	{\rm b}(\tau) = \mu_s(\theta_{\tau,N}) - \frac{d\mu_s(J)}{ds}.
	\label{eqn:biasDefinition}
\end{align}
Therefore ${\rm b}$ is only a function of $\tau$ (and not $N$) because on using an 
infinite number of samples, 
\begin{align}
	\lim_{N\to \infty}\theta_{\tau,N} = \mu_s(\theta_{\tau,N}) =  
	\int_M d\mu_s \int_0^\tau \dfrac{\partial J}{\partial u_0}\Big|_{t}
	\;\cdot\; \frac{\partial f}{\partial s}\; dt . 
	\label{eqn:expectedValueofEstimator}
\end{align}
Writing Eq. \ref{eqn:biasDefinition} more 
explicitly as,
\begin{align}
	{\rm b}(\tau) = \int_M \int_0^\tau
 	\dfrac{\partial J}{\partial u_0}\Big|_{t}
	\;\cdot\; \frac{\partial f}{\partial s}\; dt \: d\mu_s  - 
	\int_0^\infty \int_M
	\dfrac{\partial J}{\partial u_0}\Big|_{t}
	\; \cdot \; \frac{\partial f}{\partial s} \;d\mu_s \: dt,
	\label{eqn:bias}
\end{align}
clearly indicates why there is a non-zero bias for a finite value of $\tau$. 
As an optimistic estimate of ${\rm b}(\tau)$, we choose
an exponential decay at a problem-dependent rate denoted $\gamma_1$, using the 
following justification. In a uniformly hyperbolic system, a tangent vector 
can be decomposed, at every point in phase space, into its stable 
and unstable components. A stable (unstable) tangent vector 
would diminish in norm along a forward (backward) trajectory at an exponential
rate. More precisely, for almost every $u_0$, if $v^h(0,u_0)$ is a stable tangent vector at $u_0$, 
there exist $C, \alpha > 0$ such that, 
\begin{align}
	\label{eqn:stablePerturbation}
	\norm{v^h(t, u_0)} \leq C \exp{(-\alpha t)}\norm{v^h(0,u_0)},\; \text{ for all }\;t \geq 0.
\end{align}
That is, in uniformly hyperbolic systems, solving the homogeneous tangent equation 
with a stable tangent vector as the initial 
condition, results in a stable tangent vector whose norm grows smaller exponentially. It must be mentioned here 
that the interval $[-\alpha,\alpha]$ (with $\alpha$ defined by Eq. \ref{eqn:stablePerturbation}) 
lies in the gap between the largest negative and smallest positive Lyapunov exponents. 
That $\alpha$ provides a lower bound for the smallest positive Lyapunov exponent
will be used in our analysis later. Let us 
now examine the bias when the initial perturbation given by $(\partial f/\partial s)(u_0) $
is a stable tangent vector at every $u_0$. From Eq. \ref{eqn:bias},
\begin{align}
	\label{eqn:stableBias}
	{\rm b}(\tau) &= -\int_M \int_\tau^\infty \dfrac{\partial J}{\partial u_0}\Big|_t
	\;\cdot\; \frac{\partial f}{\partial s}\Big|_0 \; dt \; d \mu_s = -\int_M \int_\tau^\infty \dfrac{\partial J}{\partial u}\Big|_t
	\;\cdot\; v^h(t,u_0)\; dt\;d \mu_s \\
	\label{eqn:stableBiasCS}
	\implies \abs{{\rm b}(\tau)} &\leq  \int_M \int_\tau^\infty  \norm{\dfrac{\partial J}{\partial u}\Big|_t} 
	\;(C \exp{(-\alpha t)}) \norm{\dfrac{\partial f}{\partial s}\Big|_0} \; dt \; d \mu_s \\
	\label{eqn:stableBiasUpperBound}
	&\leq \dfrac{C}{\alpha} \;\norm{\dfrac{\partial J}{\partial u}}_\infty
 \;	\norm{\frac{\partial f}{\partial s}}_\infty \; \exp{(-\alpha \tau)}.
\end{align}
The norm $\norm{\cdot}_\infty$ of a vector field $X$ expressed in coordinates 
as $ X(u) = (X_1(u), X_2(u), \cdots, X_d(u))$ is defined as
$\norm{X}_\infty:= \sup_{u\in M} \norm{X(u)}$, where the norm of the 
tangent vector $X(u)$ is the $l^2$-norm, $\norm{X(u)}:= \sqrt{(\sum_{i=1}^d \abs{X_i(u)}^2)}$. We obtain 
Eq. \ref{eqn:stableBias} by recognizing that the order of
integration can be swapped in the second term (the true sensitivity) in Eq. \ref{eqn:bias} in this case
since the 
absolute value of the integrand is bounded for all time. Equation 
\ref{eqn:stableBiasCS} follows from using
the Cauchy-Schwarz
inequality and the fact that 
stable perturbations decay in time, as described by inequality \ref{eqn:stablePerturbation}. 
Thus, we obtain
that the bias of the estimator $\theta_{\tau, N}$ decays exponentially with $\tau$ if the perturbations
lie entirely in the stable subspace of the tangent
space at every point. In general, the initial tangent vector will also 
have an unstable component. The unstable contribution to the 
bias follows the decay of time correlations \cite{ruelle} in the system, which has 
been shown to be exponential, at best, in uniformly hyperbolic systems. 
Collet and Eckmann \cite{collet} 
proposed the conjecture that for randomly picked observables 
in expanding systems (where all the Lyapunov exponents are stricly 
positive), the decay of time correlations is at an exponential rate 
that is at least as slow as the smallest positive Lyapunov exponent. 
Applying the Collet-Eckmann conjecture provides 
another rationalization of the fact that even in the optimistic 
event of the perturbation field being stable 
(Eq.\ref{eqn:stableBias} - Eq.\ref{eqn:stableBiasUpperBound}), we 
would expect the time for the decay of the bias (and hence the minimum
time required for the convergence of the ES method) to be 
at least that of the decay of time correlations in the system. This is 
because the exponential rate at which the bias decays in the stable
perturbation case is $\alpha$ (from Eq.\ref{eqn:stableBiasUpperBound}), 
a lower bound on the smallest positive Lyapunov exponent, 
which assuming the Collet-Eckmann conjecture applies, is on the same 
order as the decay of time correlations in the system.

Estimating the decay of correlations 
is an active research area and previous studies \cite{baladi1,baladi2} have obtained 
that even among hyperbolic chaotic systems, 
\emph{intermittent} systems can exhibit subexponential decay of correlations.
Therefore, an exponential decay of the bias with 
integration time is justified as a representation of the optimal
scenario, giving rise to the following model for the squared bias, for some constant $C_{\rm b} > 0$:
\begin{align}
	{\rm b}^2(\tau) = C_{\rm b} \exp{(-2 \gamma_1 \tau)},
\end{align} where $\gamma_1$ is a problem-dependent rate. In the same vein as our discussion on 
the bias above, we propose a model
for the best case variance and provide a justification for the chosen model.
The ES estimator $\theta_{\tau,N}$ is a sample average of the 
random variable $\int_0^\tau (\partial J/\partial u_0)\Big|_t
\;\cdot\;(\partial f/\partial s)\Big|_0 \; dt$,
the randomness arising in the (deterministic) chaotic system due to the randomness in the initial 
condition. We know that the initial conditions are distributed according to $\mu_s$ 
and this gives rise to an unknown $\tau$-dependent distribution for 
$\int_0^\tau (\partial J/\partial u_0)\Big|_t
\;\cdot\;(\partial f/\partial s)\Big|_0 \; dt$. For a finite
$\tau$, we assume that the variance of this distribution is finite. 
Then, it follows that since $\mu_s(\theta_{\tau,N})$ is bounded 
as we established above, the central limit theorem (CLT) applies and therefore,
for large $N$,
\begin{align}
	{\rm var}(\theta_{\tau, N}) \to \frac{ 
	{\rm var}( \int_0^\tau (\partial J/\partial u_0)\Big|_t\;
	\cdot\;(\partial f/\partial s)\Big|_0\; dt)}{N}.
	\label{eqn:CLT}
\end{align}
The applicability of the CLT for the distribution of $\theta_{\tau,N}$ is 
in general an optimistic assumption, as discussed in previous works \cite{eyink}
and in the numerical example
in section \ref{sec:lorenz63}.
It is reasonable to expect that 
${\rm var}( \int_0^\tau (\partial J/\partial u_0)\Big|_t\;
\cdot\;	(\partial f/\partial s)\Big|_0\; dt )$ increases
exponentially with $\tau$ since for almost every initial state, 
$\abs{(\partial J/\partial u_0)\big|_t\;
\cdot\;	(\partial f/\partial s)\big|_0} \sim {\cal O}(\exp(\lambda_1 t)),$ where 
$\lambda_1$ is the largest Lyapunov exponent of the system. Therefore,
we expect that the variance grows exponentially at 
the rate of twice the largest 
Lyapunov exponent of the system. Thus, we propose the following optimistic
model for the variance, for some $C_{\rm var} > 0$,
\begin{align}
	{\rm var}(\tau, N) = \frac{C_{\rm var} \exp{(2 \lambda_1 \tau)}}{N} 
	\label{eqn:varianceModel}.
\end{align}
\subsection{Optimistic convergence estimate for the ES method}
\label{sec:model}

Using the optimistic estimates for the bias and the variance described in 
section \ref{sec:optimisticEstimates}, we arrive at the following ansatz 
for the mean squared error in the ES estimator:
\begin{align}
		\tilde{e}(\tau, T) 
		&= {\rm b}^2(\tau) + {\rm var}(\tau,N) \\
		&= C_{\rm b} \exp{(-2 \gamma_1 \tau)} + \frac{C_{\rm var} \tau}{T} \exp{(2 \lambda_1 \tau)},	
\label{eqn:err}
\end{align} where we use $T := N \tau$ to denote the computational cost.
The relationship between the integration time $\tau^*$ that minimizes the 
mean squared error and the cost $T$ for the model in Eq. \ref{eqn:err} is:
\begin{align}
	\tau^*(T) = \argmin_{\tau} \tilde{e}(\tau, T) 
	=  -\frac{1}{2 \lambda_1} + \frac{{\cal W}(c)}{2 (\gamma_1 + \lambda_1)}
		\label{eqn:mintau}
\end{align} where, $$c = 2 \; C_{\rm b}\frac{ \gamma_1 (\gamma_1 + \lambda_1)T}{C_{\rm var} \lambda_1}\; \exp{(1 + \gamma_1/\lambda_1)}$$ and ${\cal W}$ is the Lambert $W$-function.
The above relationship shows that, given constants $C_{\rm b}$
and $C_{\rm var}$ independent of $\tau$, the 
optimal trajectory length of each independent sensitivity
evaluation, $\tau^*$, varies sub-logarithmically with the cost $T$. 
\begin{figure}
%\begin{multicols}{2}
	\includegraphics[width=\linewidth]{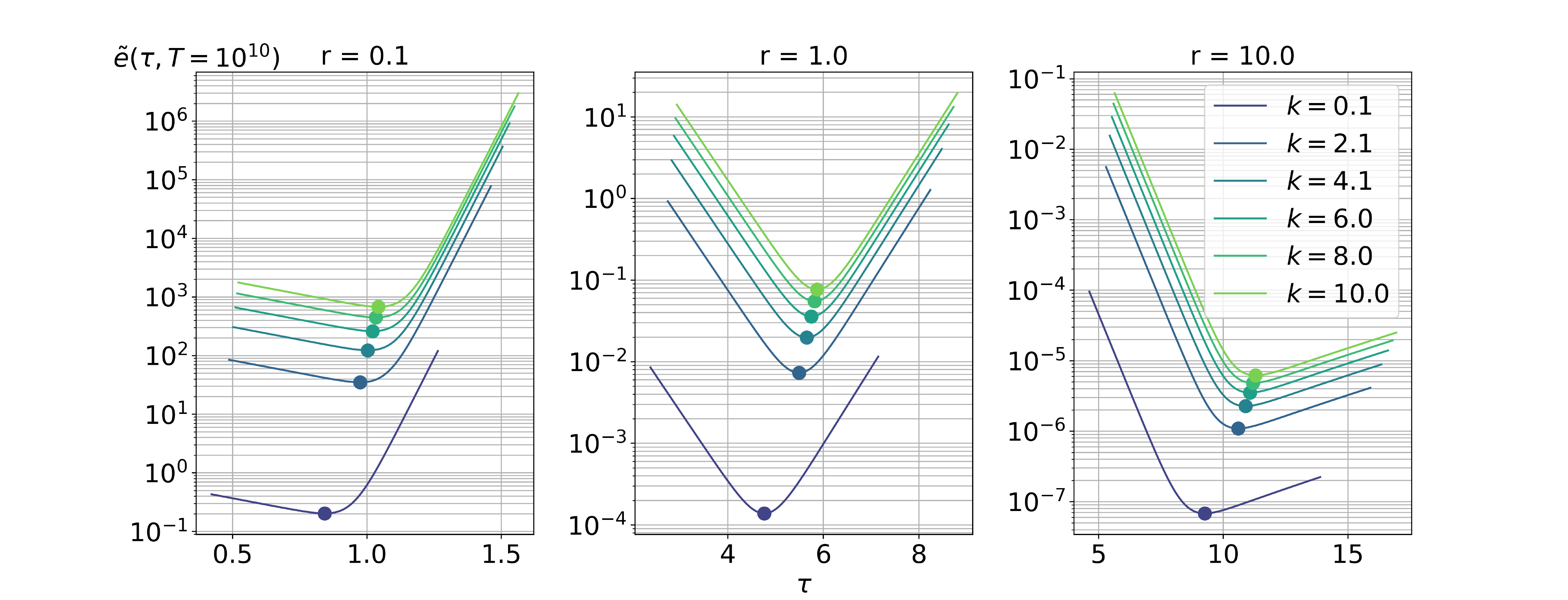}\par 
		\caption{The mean squared error as a function of integration
		time $\tau$, for different values of the variables $k$ and $r$. 
		The optimal value $\tau^*$ is marked on each of the 
		plots.}
		\label{fig:evstau}
\end{figure}

\begin{comment}
\begin{itemize}
				\item change plot legend - do not introduce $k_0$ and $r_0$, they are 
non-standard.
	\item decay of correlations -> decay of bias.

	\item in the $k^*$ plot, say what the value of $r^*$ is.
	\item exponent of power law is controlled by $r$ and the coefficient 
			of the power law which is manifested as the height in figures
						1 and 2 is controlled by $k$.

	\item the factor of increase in the cost to cause a proportional decrease
			in the error is controlled by the exponent of the power law. 
		
	\item give analysis after factor plot. $\gamma_1$ is the slowest
			timescale that contributes to the decay of bias. when 
						$\lambda_1/\gamma_1 = 1$, we need ~ 6 times ...
						when the ratio is 10, we need ...

	\item Run for more samples, get less noisier plot.

	\item Get difference plot showing difference between perturbed and unper
			turbed vane problem from Chai's defence.

	\item add grdi to factor cost plot
		\end{itemize}
	\end{comment}

\begin{figure}
	\begin{subfigure}{0.5\textwidth}
		\includegraphics[width=\textwidth]{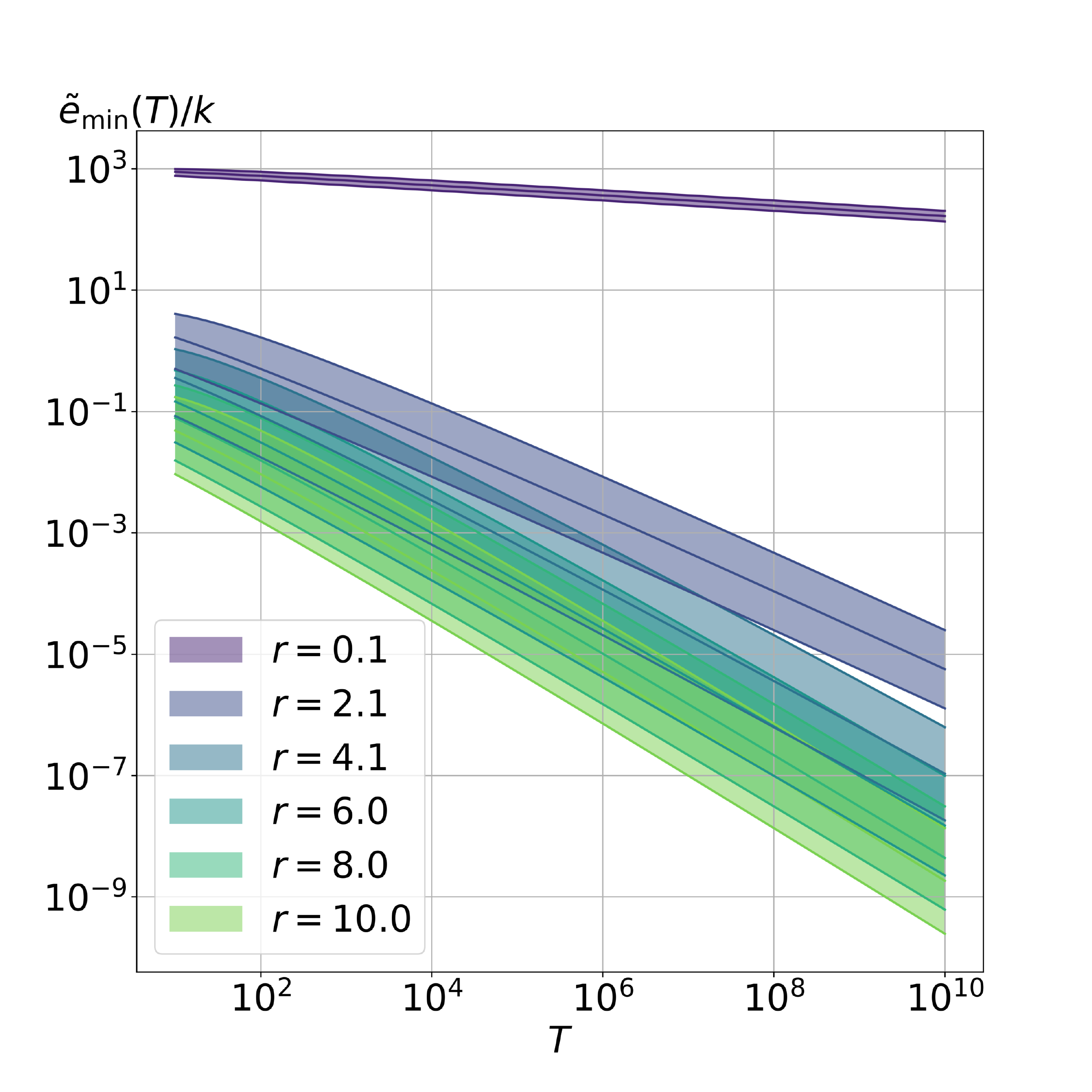}\par 
		\caption{$\tilde{e}_{\rm min}/k$ as a function of $T$,
		at different values of $r$.}
		\label{fig:eminbykvsr}
\end{subfigure}
	\begin{subfigure}{0.5\textwidth}
		\includegraphics[width=\textwidth]{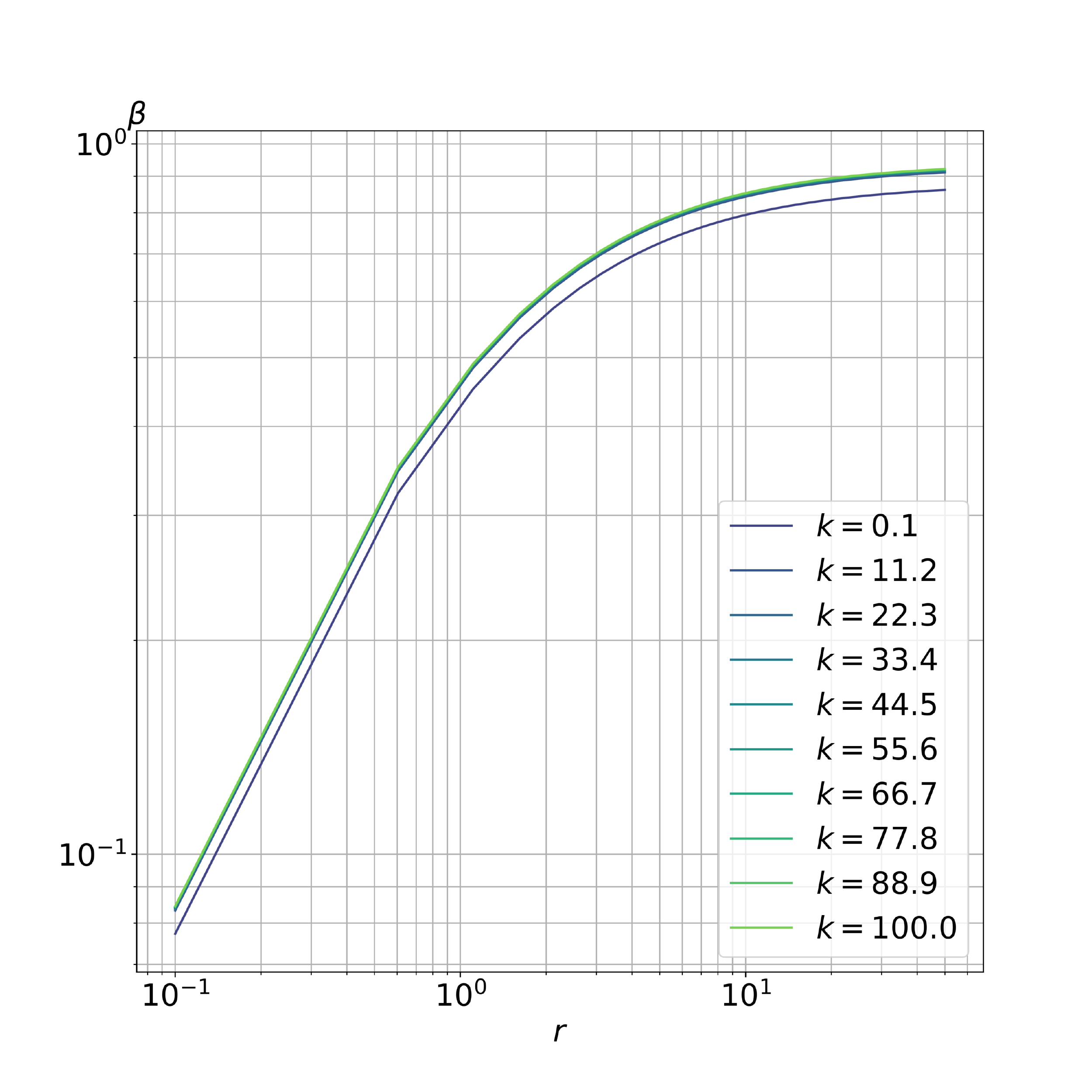}\par
\caption{Rate of convergence as a function of  
	the ratio $r = \gamma_1/\lambda_1$.}
\label{fig:rateofConvergence}
\end{subfigure}
\end{figure}

From Eqs. \ref{eqn:err} and \ref{eqn:mintau}, it can be 
seen that the least mean squared error, denoted by 
$\tilde{e}_{\rm min}(T) := \tilde{e}(\tau^*(T), T)$ can be reduced to 
a function of the variables, $k := C_{\rm b} \gamma_1/C_{\rm var}$ and 
$r := \gamma_1/\lambda_1$. Figure \ref{fig:evstau} shows the variation 
of $\tilde{e}(\tau, T)$ with $\tau$ at 
a fixed $T$, for different values of $k$ and $r$. 
It is clear that the mean squared error is high in magnitude when 
$r < 1$ no matter the trajectory length, $\tau$. As expected for $r>1$, the optimal trajectory 
length increases with $r$ and leads to smaller 
mean squared errors. The mean squared error is large,
as expected, both when a) $\tau$ is so short that 
variance is negligible, and the bias is significant
and b) $\tau$ is large enough that the effect of exponential 
increase in the variance is clear. Thus we see a roughly parabolic 
shape centered at the optimal trajectory length for $r=1$,
where the competing timescales are equal. The parabola is 
	asymmetric $r<1$ because the time constant of the squared bias decay $(1/(2\gamma_1))$
is larger than that for the variance increase ($1/(2\lambda_1)$).
Thus, the bias drops more slowly when compared to $r\geq1$,
and upon increasing $\tau$, is overwhelmed very quickly by the 
dramatic increase in the variance, which is reflected in  
${\tilde{e}(\tau,T)}$ for $\tau > \tau^*$. Analogously, the downward deflection 
of the right arm of the parabola for $r>1$, is explained by the 
fact that for values of $\tau > \tau^*$, we have a negligible 
bias and the exponential increase of the variance is slow, when 
compared to at $r=1$, while for $\tau < \tau^*$, increasing
the trajectory length dramatically drops the bias but the variance
remains small, and hence presents
as a quick decline in the mean squared error. In \ref{fig:eminbykvsr}, $\tilde{e}_{\rm min}(T)/k$ is shown against the total 
computational cost $T$ at different values of $r$. The plots in \ref{fig:eminbykvsr} reveal
that $\tilde{e}_{\rm min}$ is shifted upward on decreasing $k$ but the slope of $\tilde{e}_{\rm min}$ 
vs. T, is relatively unaffected by a change in $k$ when compared to a change in $r$. 
From \ref{fig:eminbykvsr}, we also observe that the least mean squared error  $\tilde{e}_{\rm min}$ 
decays like an approximate power law of the total cost $T$.
That is,
	\begin{align}
		\tilde{e}_{\rm min} \sim {\cal O}(T^{-\beta})\;\;
\text{ for some } \beta \equiv \beta(k,r) > 0. 
\label{eqn:beta}
\end{align}

In \ref{fig:rateofConvergence}, the rate of convergence $\beta$ is reported 
at different values of $r$ and $k$. From \ref{fig:rateofConvergence}, 
$\beta$ appears to be quite robust to varying
the ratio of the bias to variance coefficients, $C_{\rm b}/C_{\rm var}$, when 
$\gamma_1$ is kept constant. On the other hand, it can be seen that the influence
of the ratio of timescales $r$ is significant on the rate of convergence. 
For values of $r > 1$, the least mean squared error falls 
faster than $1/\sqrt{T}$, for 
a range of values of $k$, indicating that convergence rates better than 
a typical Monte Carlo sampling can be achieved on choosing an optimal 
$\tau$ under the assumptions of section \ref{sec:optimisticEstimates}.
This implies that, assuming a strongly chaotic system satisfying our optmistic
estimates, when the timescale of the decay of bias is shorter than that of the 
growth of perturbations (1/$\lambda_1$), the ES method can be very efficient.
On the contrary, \ref{fig:rateofConvergence} also indicates that the number of samples required
to half $\tilde{e}_{\rm min}$ at $r = 1$ must be increased four-fold while at 
$r = 0.1$, a factor of $2^{10}$ increase in the number of samples is required to half the error. 
To summarize, even in the ideal case of exponential decay of the bias, a rate of decay
smaller than the leading Lyapunov exponent, would lead to 
a significantly less efficient method than a typical Monte Carlo.

\section{Numerical examples}
\label{sec:results}
The previous section was dedicated to a mathematical analysis of the convergence 
of the ES method. We were able to predict the best possible rates of convergence under suitable
assumptions on the dynamics. In this section, we treat numerical examples of low-dimensional
chaotic systems as well as simulations of turbulent flow. In each of the examples, 
our goal is to gauge, based on our numerical results, the applicability of 
the assumptions adopted in our analysis in section \ref{sec:analysis}. When deemed 
applicable, we estimate the rate of convergence using our results from section \ref{sec:analysis}
and discuss the computational tractability of the ES method given this rate. In other 
cases where either the analysis is inapplicable or the estimation of bias and variance 
is not practical, we use a physics-informed approach to predict the rate of convergence.
The discussions following the numerical results delineate guidelines for 
determining the practicality of the ES method. 

\subsection{The Lorenz'63 attractor}

\label{sec:lorenz63}
As noted in the introduction, Eyink \emph{et al.} \cite{eyink} 
have performed a numerical analysis 
of the ensemble adjoint and related methods on the Lorenz'63 system and make
several important observations regarding the convergence trends of $\theta_{\tau,N}$
in $\tau$ and in $N$. We choose the Lorenz'63 system as the first example in order to validate our present results against Eyink \emph{et al.}'s. 
The Lorenz'63 system is a 3-variable model of fluid convection that is used
as a classic example of chaos \cite{lorenz63}. It consists of the following
system of ODEs:
\begin{align}
\notag
\frac{dx}{dt} &= -\sigma x + \sigma y \\
\notag
\frac{dy}{dt} &= - x z + s x - y \\ 
\label{eqn:lorenz63} 
\frac{dz}{dt} &= xy - bz, 
\end{align} with the standard values of $\sigma = 10$, $b = 8/3$ and $s = 28$. 
These equations were derived by Lorenz and Saltzman \cite{lorenz63}
from the conservation equations for a fluid 
between horizontal plates maintained at a constant temperature difference.
The phase vector $u := [x, y, z]^T$, whose evolution these equations
describe, corresponds to coefficients in the Fourier series expansion 
of the stream function and the temperature profile. The parameter $s$
is the Rayleigh coefficient normalized by the critical value above 
which flow instabilities develop. The objective function of our interest
is $J(u(t)) := z(t)$ which is proportional to the deviation of the temperature 
from the linear profile that would be achieved if the fluid 
were static. 
It is well-known that a compact attractor exists in phase space. 

We use the algorithms described in section \ref{sec:algorithm} to compute
all three types of ES estimators $\theta_{\tau, N}^{\rm FD}$, $\theta_{\tau, N}^{\rm A}$ 
and $\theta_{\tau, N}^{\rm T}$. To ensure that the initial condition is sampled
from the steady-state distribution on the Lorenz attractor, the system is evolved for a
spin-up time of about $1.1$ time units, 
before we start computing the sensitivities. This spin-up time is estimated
as $1/\lambda_1$, with $\lambda_1 \approx 0.9$ known from the literature to be
the largest Lyapunov exponent. The true
value of the sensitivity $(d\mu_s(J)/ds)$ was computed by Eyink \emph{et al.} \cite{eyink} to be $\approx 0.96$. This value is obtained by numerically computing $\mu_s( J)$ as an ergodic average at a range of values of $s$. It can be seen that $\mu_s( J)$ 
turns out approximately to be a straight line with a slope of 0.96 
around $s = 28$. The primal, the tangent and adjoint dynamics in equations \ref{eqn:lorenz63}, \ref{eqn:TangentSystem} and \ref{eqn:AdjointSystem} respectively are computed using forward Euler time integration with a timestep of $0.005$ time units. The computational 
cost $T = N \tau$ was chosen to be 5000 time units. The estimators were computed for 
a range of values of $\tau$ up to 3 time units. In order to apply our analysis in section
\ref{sec:analysis}, we wish to numerically estimate the bias and variance of $\theta_{\tau,N}$. We estimate $\mu_s(\theta_{\tau,N})$ as a sample average of $\theta_{\tau,N}$
using 5 million independent samples. That is, we approximate 
$\mu_s(\theta_{\tau,N})$ as $\theta_{\tau, 5\times 10^6}$ while the true value
is not a function of the number of samples but only of $\tau$. The variance
of $\theta_{\tau,N}$ is again approximated as a sample average wherein $\mu_s
(\theta_{\tau,N})$ is replaced with its estimate $\theta_{\tau,5\times 10^6}$. 

\subsubsection{Empirically determined upper bound for rate of convergence}
Our overarching goal is to predict the rate of convergence of the ES method. 
We now discuss that it is possible to roughly estimate the rate based on 
numerical results at a fixed computational cost $T$. In Lea-Allen-Haine's 
application of the ES method on the Lorenz'63 system \cite{lea}, 
$\tau = 1$ is used to obtain an accurate estimate of  
$d\mu_s(J)/ds$; for the sake of comparison with their results, 
we estimate the rate of convergence locally around $\tau = 1$. To obtain
the rate of convergence, we obtain estimates, for the exponential rates 
of increase and decrease of the variance and the bias respectively, 
from our numerical results. Then, we use \ref{fig:rateofConvergence} 
to obtain the rate of convergence from 
the ratio of these obtained rates.

\begin{figure}
	\begin{subfigure}{0.5\textwidth}
			\includegraphics[width=\textwidth,height=8cm]{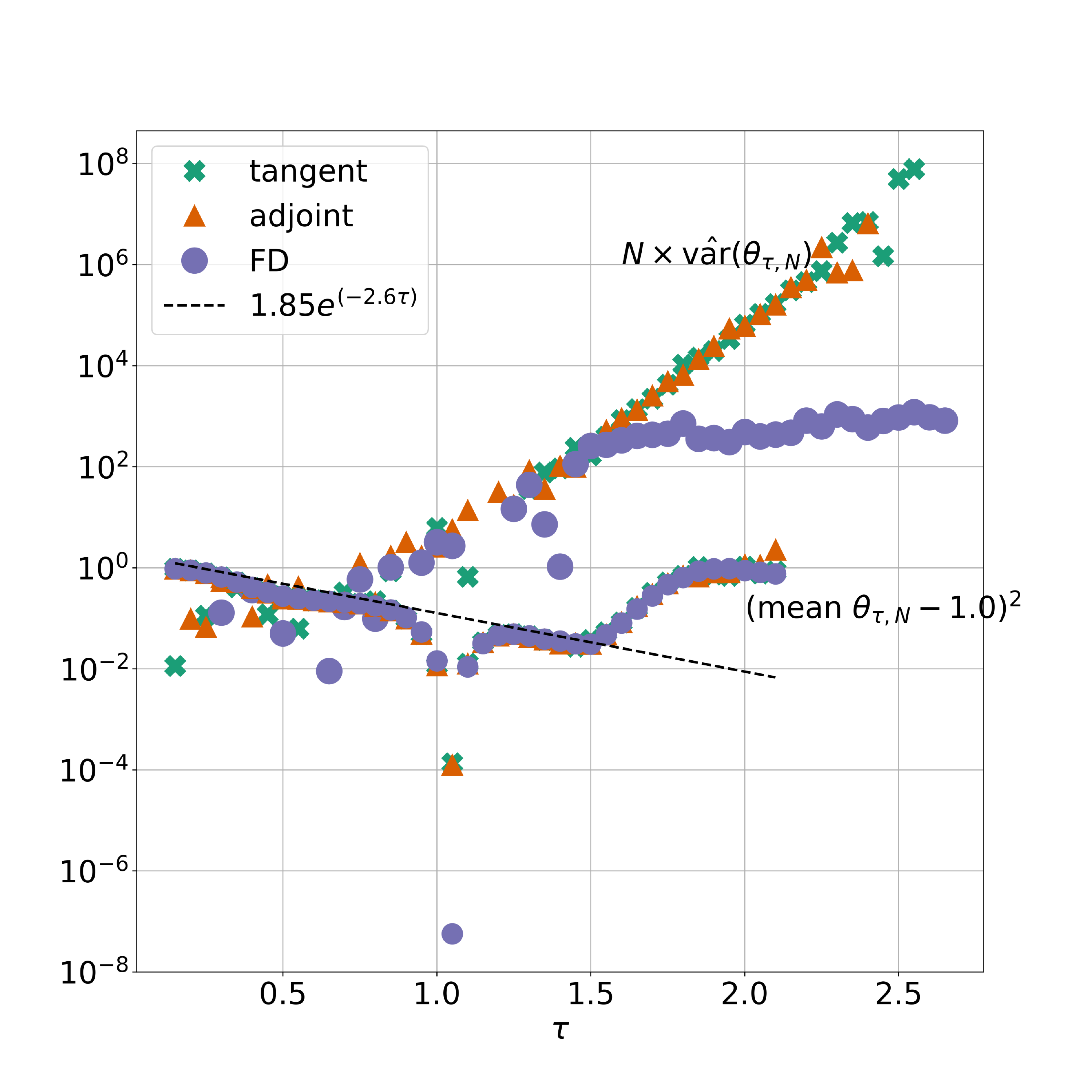}
		\caption{Estimates of the variance and the bias of $\theta_{\tau,N}$ as a function of $\tau$, for the Lorenz'63 system 
		outlined in section \ref{sec:lorenz63}. 
		The dashed line indicates the least-squares fit over $\tau \leq 1.5$. }
	\label{fig:lorenz63_bv}
	\end{subfigure}
\begin{subfigure}{0.5\textwidth}
			\includegraphics[width=\textwidth,height=8cm]{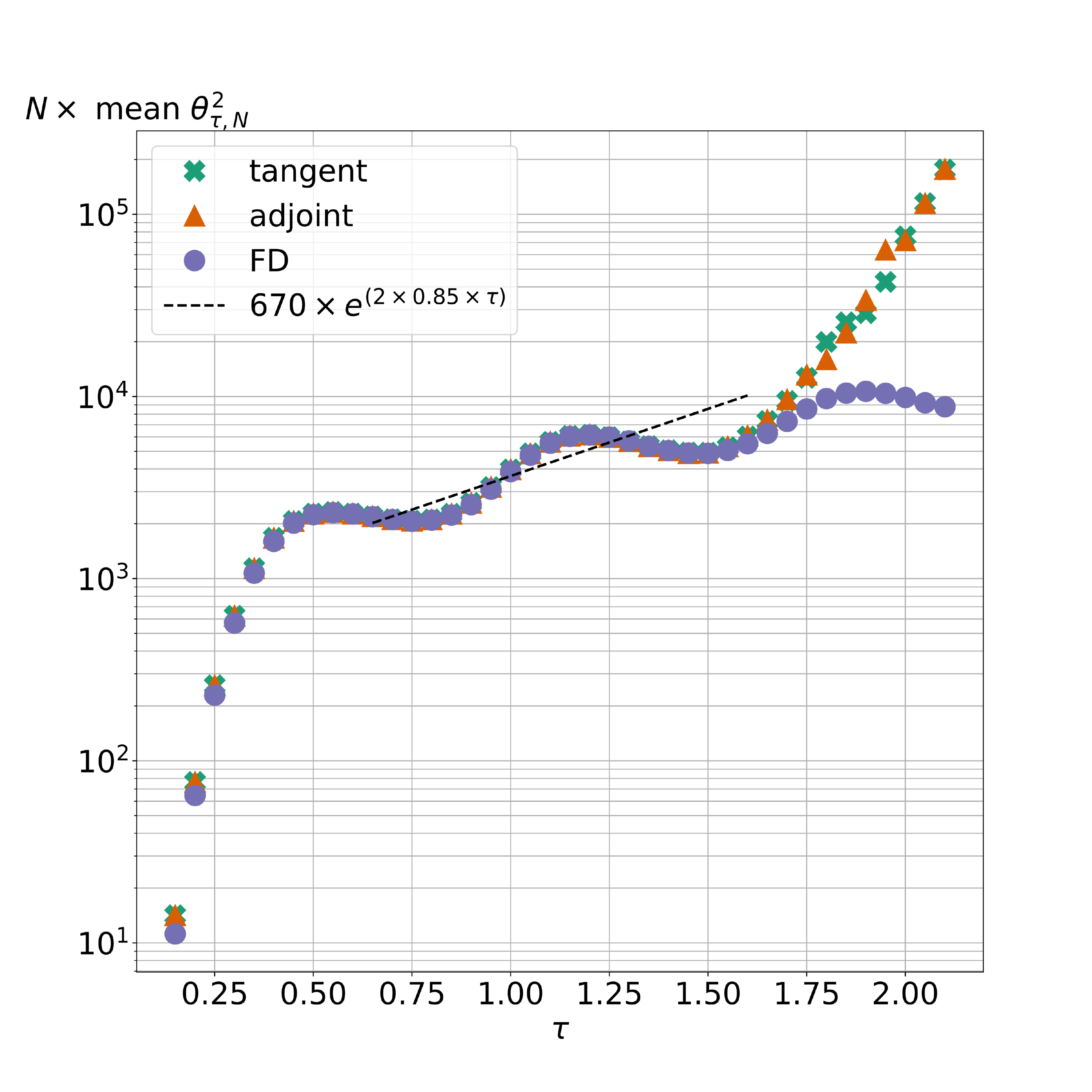}
	
	\caption{Sample mean estimates of $\mu_s(\theta_{\tau,N}^2)$ as a function of $\tau$ for the Lorenz'63 system 
	outlined in section \ref{sec:lorenz63}. 
		The dashed line indicates the least-squares fit over $\tau \leq 1.5$.}
	\label{fig:lorenz63_v}
\end{subfigure}
\end{figure}

First we establish an empirical estimate for the variance rate. Our numerical estimates 
for the bias and variance of the estimators $\theta_{\tau,N}^{\rm A}, 
\theta_{\tau,N}^{\rm T}$ and $\theta_{\tau,N}^{\rm FD}$ computed as described 
above, are shown in  \ref{fig:lorenz63_bv}. In  \ref{fig:lorenz63_v}, 
the numerical estimates of $\mu_s(\theta_{\tau,N}^2)$ are shown for all three estimators.
Ignoring transient behavior for $\tau$ up to $\sim 0.5$, the asymptotic exponential 
rate of the variance estimate, obtained by determining 
the slope from  \ref{fig:lorenz63_bv}, is larger than 
the $2 \lambda_1$ rate that we predicted in \ref{sec:model}. 
To understand this result, let us assert that our numerical estimate
of the variance reflects the trend of the true variance and the CLT holds
for the given range of values of $\tau$. If the first assertion holds, 
the true variance is increasing exponentially at a faster rate than $2\lambda_1$. 
Now, owing to the convergence of Ruelle's formula, we know that 
$\mu_s(\theta_{\tau,N})$ cannot have an unbounded growth as a function of $\tau$ 
and therefore, the rate of increase of ${\rm var}(\theta_{\tau,N})$ must be 
captured by that of $\mu_s(\theta_{\tau,N}^2)$. Computing the slope from
 \ref{fig:lorenz63_v}, we see that the numerical estimate of 
$\mu_s(\theta_{\tau,N}^2)$ is indeed exponential
in $\tau$ at the rate $\sim 2 \times 0.85$ for both
the tangent and adjoint estimators, from a least-squares
fit. This value of the rate is closer to our
theoretical prediction of $2 \lambda_1$. Therefore, it is reasonable to conclude
that neither of our previously stated assertions holds true. As a result, we 
have considerable error in our estimates of $\mu_s(\theta_{\tau,N})$ 
and $\mu_s(\theta_{\tau,N}^2)$ since the error is decaying slower than expected from CLT. 
 Since both these errors play a role in 
the variance estimation, obtaining the rate of increase from the 
estimate of $\mu_s(\theta_{\tau,N}^2)$ is more accurate.
It is thus reasonable to take the better numerical estimate, 
$2 \times 0.85$, as the rate of exponential increase of the variance for $\tau$ up to 1.5.

In \ref{fig:lorenz63_bv}, we obtain from the slope of the bias term (again using a least-squares fit) 
for $\tau \leq 1.5$, the rate $\gamma_1 \sim 1.3$. 
Therefore, we obtain $r = 1.3/0.85 = 1.5$ as the rough estimate needed to determine
the rate of convergence under our model assumptions in sec \ref{sec:model}.
Thus, we note from  \ref{fig:rateofConvergence} that $\beta \sim  0.5$. It would not be gainful to 
also estimate $k$ since we only seek an upper bound for $\beta$ which is quite 
insensitive to $k$, in the first place. We can interpret
this rate as the best possible rate of convergence for the Lorenz'63 system.

\subsubsection{Agreement with Eyink \emph{et al.}'s results}
Eyink \emph{et al.} \cite{eyink} propose that the probability distribution 
of $\theta_{\tau,N}$ for the Lorenz'63 system is a fat-tailed distribution and 
does not obey the classical CLT. Consequently, our assumption of 
bounded variance of $\theta_{\tau,N}$ for finite values of $\tau$ should
fail for this system. Our numerical results indicate that both 
the variance and bias trends are worse than our optimistic model, thus
confirming their predictions. However, locally around $\tau \sim 1$, 
operating under the assumption 
that the bias and the variance trends can be empirically modelled using
our optimistic estimates, we were able to provide an upper bound on the actual convergence
rate. Owing to the fact that $\tau \sim 1$ is not large enough for the failure of the CLT
assumption to be manifest, our empirically determined rate of 0.5 provides an 
upper bound on that estimated using Eyink \emph{et al.}'s tail estimates. 
Thus we conclude firstly that our numerical results provide further evidence 
in support of the failure of the CLT for $\theta_{\tau,N}$. Secondly, 
empirical determination
of the optimistic rate of convergence under the CLT assumption locally around 
$\tau \sim 1$, predicts an upper bound at the rate of a Monte Carlo simulation, 
confirming Eyink \emph{et al.}'s observations.

It is worth commenting on the fundamental difference between the 
distribution of sample averages of state functions and that of 
$\theta_{\tau,N}$, which is a sample average of the derivative
of a state function (averaged over a short time). The former distribution has bounded variance but the latter
is not guaranteed to, as this example demonstrates. In fact, it has been shown
\cite{lorenz63_CLT} that $N$-sample averages (computed numerically as ergodic 
averages) of state functions converge at the rate of $1/\sqrt{N}$. However, the
behavior of the $N$-sample average that is $\theta_{\tau,N}$, is unlike that of 
state functions. $\theta_{\tau,N}$ converges to $\mu_s(\theta_{\tau,N})$ 
(due to the law of large numbers) at a rate slower than $1/\sqrt{N}$. 
This rate decreases as $\tau$ increases, bearing on the poor accuracy of the estimate 
$\theta_{\tau,5\times 10^6}$ for values of $\tau \gtrsim 1.5$ seen
in  \ref{fig:lorenz63_bv}, despite using a seemingly large number of samples.

Another observation that can be made from  \ref{fig:lorenz63_bv} is
that unlike the variances of $\theta_{\tau,N}^{\rm A}$ and 
$\theta_{\tau,N}^{\rm T}$, ${\rm var}(\theta_{\tau,N}^{\rm FD})$ appears
to saturate for $\tau \gtrsim 1.5$. This is explained by the fact 
that $\abs{\theta_{\tau,N}^{\rm FD}}$ is bounded above by $\norm{J}_\infty/\epsilon = c/\epsilon$, 
where $\epsilon$ is the value of the parameter perturbation in the finite
difference approximation. The value $c$ is the supremum of the $z$ coordinate
of the Lorenz attractor, which is a finite value since the attractor is a bounded set. The fact that 
$\theta_{\tau,N}^{\rm FD}$ is bounded
for all $\tau$ can also be observed in the saturation of the estimate
of $\mu_s(\theta_{\tau,N}^{\rm FD^2})$ shown in  \ref{fig:lorenz63_v}. 

\subsection{The Lorenz'96 model}
\label{sec:lorenz96}
\begin{figure}
		\includegraphics[width=0.7\textwidth]{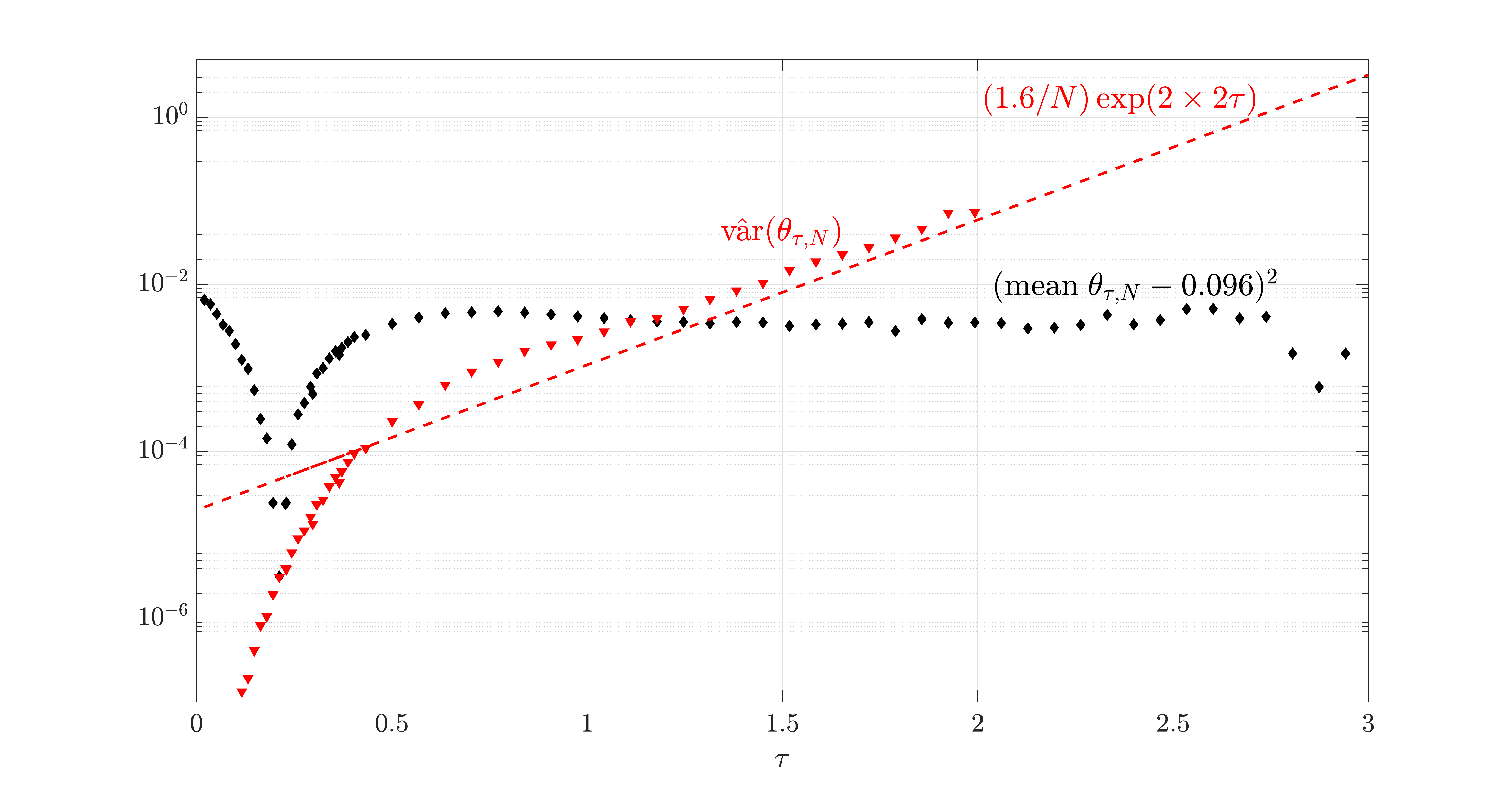}
		\caption{The variance and the 
square of the bias in the ensemble tangent sensitivity estimator
		for the Lorenz'96 system. The dashed 
		line indicates the least-squares fit of the variance
		data.}
		\label{fig:lorenz96}
\end{figure}

Our results for the Lorenz'63 system in section \ref{sec:lorenz63} 
show that an upper bound on the rate of convergence for an 
optimal choice of $\tau$ ($\sim 1$) is $0.5$. Although the 
actual rate is slower
than a typical Monte Carlo simulation, one can immediately see that since the 
system is low-dimensional, ES is still practical in the Lorenz'63 system. 
In this section, we aim to assess the computational practicality of the ES
method in a higher-dimensional system. We consider the 
atomspheric convection model, called the Lorenz'96 \cite{lorenz96} model,
that is known to have a strange attractor \cite{karimi}. The 40-dimensional
state vector $u := [x_1,\cdots,x_{40}]^T$ evolves according to the 
following set of odes \cite{lorenz96,karimi}
\begin{align}
		\frac{d x_k}{dt} = (x_{k+1} - x_{k-2}) x_{k-1}  - x_k + s, \;\;
		k=1,\cdots, 40,
		\label{eqn:lorenz96}
\end{align}
where, $s$, the parameter of interest, 
denotes an external driving force. The first term 
on the right hand side of equation \ref{eqn:lorenz96} 
represents nonlinear advection
and the second term represents a viscous damping force. 
The components of the state vector 
are periodic in the sense that $x_{k} = x_{40+k},\; \;
k \in \mathbb{N}$. We define our objective function $J(u(t))$ to be
the mean of the components of $u$, i.e.,
$$
		J(u(t)) = (1/40) \sum_{k=1}^{40} x_k(t).
$$
We use the value $s=8.0$ for the forcing term, at which the 
system has been shown to exhibit chaotic behavior \cite{karimi}.
Time integration of the primal system of equations \ref{eqn:lorenz96} 
is performed using a fourth order Runge-Kutta scheme with a timestep of 
0.01 time units. The distribution of each component of the state vector
for this system is known from the literature \cite{mori} 
to converge to a Gaussian and 
the time average of $J$ approximately becomes a linearly varying function with $s$,
on long-time evolution. The slope of $\mu_s( J )$ vs. $s$ estimated 
from our computations and previous work \cite{karimi} is $0.096$.

In Fig.  \ref{fig:lorenz96}, we report the squared bias and the variance 
of the estimator $\theta_{\tau,N}^{\rm T}$ as a function of $\tau$.
The total cost, $T$, is set at $10^4$ time 
units across different values of $\tau$. 
The values of $\mu_s(\theta_{\tau,N})$ and $\mu_s(\theta_{\tau,N}^2)$ 
are computed as sample means and used in the approximation of the 
bias and variance (denoted by $\hat{\rm var}(\theta_{\tau,N})$ in Fig.  
\ref{fig:lorenz96} to indicate
the approximation as a sample mean) terms, 
identical to our description
in section \ref{sec:lorenz63} for the Lorenz'63 system. 
From Fig.  \ref{fig:lorenz96}, we can see that the
variance follows our model assumptions in section \ref{sec:optimisticEstimates}; 
we find that the variance is exponential in $\tau$
at the rate $\approx 4$, 
which is close to twice the leading Lyapunov exponent
reported in the literature \cite{karimi,mori}. 
From Fig.  \ref{fig:lorenz96}, ignoring initial
transients, it can be
seen that the bias term does not show an exponential convergence, not
even locally in the vicinity of $1/\lambda_1 \approx 0.5$. 
As a result, our local analysis
to produce a rough estimate under our model assumptions, 
as we did in section \ref{sec:lorenz63}, is not applicable in this case.
This implies that the rate of convergence is worse than
our model in \ref{sec:optimisticEstimates} (for any $r$) since the 
bias falls slower than the assumed exponential. One can argue that the mean squared error (sum of the squared 
bias and the variance) is low
in absolute value for $\tau$ near 0.5 deeming the ES estimator 
to be reasonably accurate and therefore practically applicable, 
although the asymptotic rate of convergence in $\tau$ may be low. However, note that the rate 
of convergence is an objective function-independent measure 
of the ES estimator -- the 
mean squared error may be coincidentally within required accuracy
for a choice of $\tau$, for this particular objective function.
Thus, it is reasonable to conclude that the ES method is infeasible
for the Lorenz'96 model.

\subsection{Chaotic flow over an airfoil}
\label{sec:airfoil}

%{\color{red}{To-do: {\it Airfoil location}, {\it Entropy wave due to parameter perturbation}, arrow for entropy wave parallel to freestream velocity)}}.

%{\color{red}{To-do: In this section, we talk about EA (see in red below). Do we want to change that to ET or something like that?}}

In this example, we discuss the numerical simulation of an unsteady, chaotic flow around a two-dimensional airfoil. So far, we have
assessed the convergence of the ES method by observing the trends in the bias and variance of $\theta_{\tau,N}$ in low-dimensional
systems. Our numerical results were informative enough to predict the rate of 
convergence while simultaneously being within the limits of practical computation, owing to the low dimensionality
of the systems considered in sections \ref{sec:lorenz63} and \ref{sec:lorenz96}. In contrast, in 
a typical chaotic CFD simulation, it would not be practical to numerically estimate the bias and variance 
trends of $\theta_{\tau,N}$. We will thus attempt to predict the convergence 
trend using a single finite-difference solution (that can be used to compute one sample of 
$\theta_{\tau,N}$ ). Our goal is to use a physics-based approach  
that eliminates the need for a rich $\theta_{\tau,N}$ dataset. We consider the NACA 0012 airfoil at the 
Reynolds number $Re_{\infty} = 2400$ and Mach number $M_{\infty} = 0.2$ at an angle of attack $\alpha = 20^\circ$. 
Although the flow physics in three-dimensional turbulent flows is more complex, the two-dimensional airfoil case we consider exhibits the phenomena of stall and flow separation that are responsible for the chaotic behavior. For an extensive analysis of 
the Lyapunov spectrum and its dependence on the numerical discretization for this problem, see \cite{pablo1,pulliam}.     
\begin{figure}
%\begin{subfigure}{0.5\textwidth}
		\includegraphics[width=0.5\textwidth]{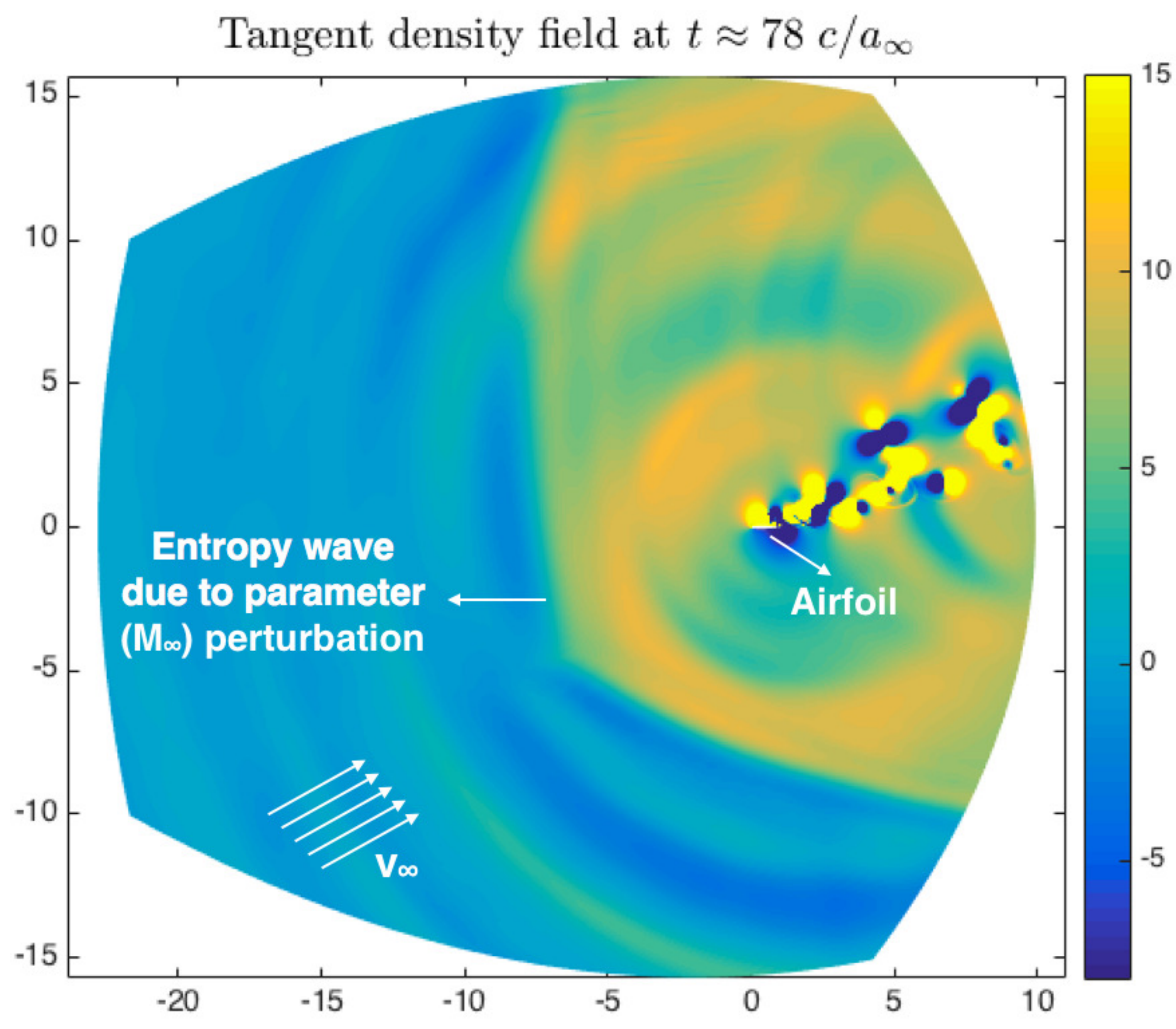}
	\caption{Tangent density field $v_{\rho}$ at time $t = 78 \, c / a_{\infty}$.} 
			\label{fig:airfoil_colormap}
\end{figure}
\begin{figure}
		\centering
		\includegraphics[width=\textwidth]{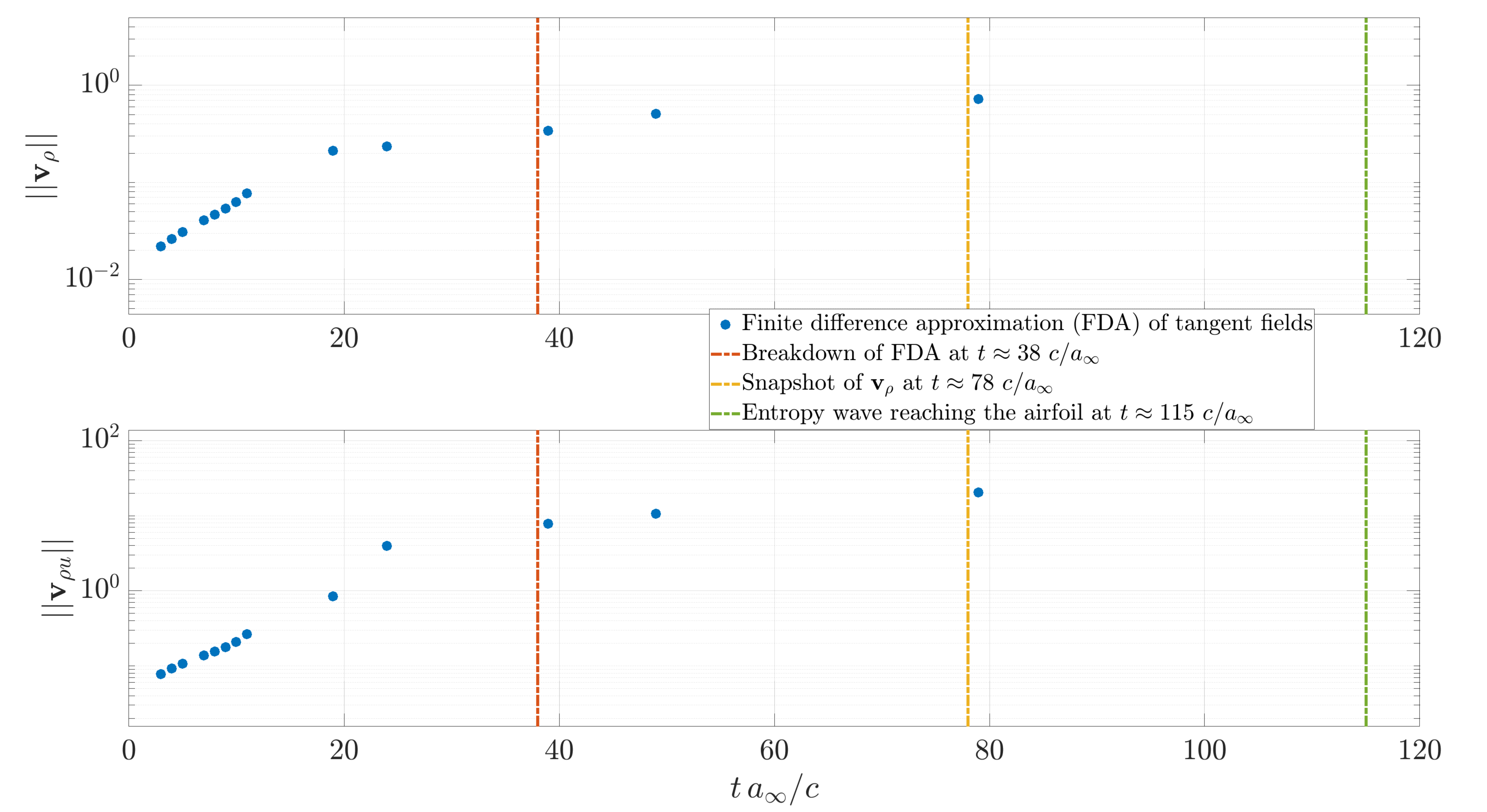}
		\caption{Time evolution of $L^2$ norm of the tangent fields corresponding to the density ${\rho}$ (top) and $x$-momentum ${\rho u}$ (bottom).}
	\label{fig:airfoil}
\end{figure}
The primal system is the set of compressible Navier-Stokes equations, which is discretized in space using a third-order Hybridizable Discontinuous Galerkin (HDG) method \cite{pablo2,pablo3} with the Lax-Friedrichs-type stablization matrix in \cite{Fernandez:2017:AIAA:SGS}. The computational domain spans $\sim 10-20$ chord lengths away from the airfoil and is partitioned using an O-grid with 35280 isoparametric triangular elements. 
The use of a large computational domain is customary in external aerodynamics to reduce the change in the effective angle of attack induced by the missing vortex upwash.
The state vector $u$ is defined to be the set of coefficients in the basis function representation 
of the density, the components of the momenta and the energy, at all elements in the computational domain. 
Our objective function $J$ is the lift coefficient and the parameter of interest $s$ is the freestream Mach number, $M_\infty$.
A three-stage, third-order diagonally-implicit Runge-Kutta (DIRK) method \cite{Alexander:77} is used for the temporal discretization. 
A no-slip, adiabatic wall boundary condition is imposed on 
the airfoil surface, and the characteristics-based, non-reflecting 
boundary condition in \cite{pablo2,pablo3} is used on the outer boundary. 
The spin-up time before the sensitivity computation is performed is 
$t = 10,000 \, c / a_{\infty}$, where $a_{\infty}$ denotes the far-field speed 
of sound and $c$ is the chord length. We note that the chosen spin-up time 
is one order of magnitude larger than the time required for 
the convergence of time-averaged flow quantities \cite{pablo1}. 
At the spin-up time, we reset $t = 0$ and all the results below are indicated with respect to this new initial time.
%and a third-order diagonally-implicit Runge-Kutta DIRK. In particular, third-order HDG and DIRKThe HDG solver for the primal uses a third order accurate scheme in space and a third order accurate diagonally implicit Runge-Kutta scheme for time integration.
%that spans 10 chord lengths away from the airfoil in all directions and

Since a tangent solver is not available typically in CFD codes, the 
tangent perturbation fields denoted by $v(t)$ are computed non-intrusively with regard to the primal solver using a
finite difference approximation (FDA). The parameter perturbation has a magnitude of $\epsilon = 10^{-6}$. 
Fig. \ref{fig:airfoil_colormap} shows the sensitivity field of the fluid density 
with respect to the perturbation in the freestream Mach number 
at time $t = 78 \, c / a_{\infty}$. 
This quantity, denoted in Fig. \ref{fig:airfoil} as $v_{\rho}$, can be 
used to compute a single sample of $\theta_{\tau,N}^{\rm FD}$. The airfoil, much smaller in size than the computational 
domain, is annotated in Fig.  \ref{fig:airfoil_colormap}. Entropy and acoustic waves are generated on the outer boundary at $t = 0$ due to the $M_{\infty}$ perturbation. From the entropy wavefront shown in Fig.  \ref{fig:airfoil_colormap}, one observes that 
the entropy wave does not reach the airfoil yet at $t = 78 c / a_{\infty}$. The time evolution of the $L^2$ norm of the tangent fields corresponding to the density, $v_{\rho}$, and the $x$-momentum, $v_{\rho u}$, are shown in \ref{fig:airfoil}. As we discussed in section 
\ref{sec:lorenz63}, finite difference sensitivities tend to saturate with time
as opposed to tangent and adjoint sensitivities which keep 
increasing exponentially. This is due to the finite difference sensitivities having an upper bound proportional to 
$1/\epsilon$, since the supremum of the objective function over phase space is a finite value independent of time. As indicated in Fig.  \ref{fig:airfoil}, 
the entropy wave reaches the airfoil at $t \sim 115 \, c / a_{\infty}$. By this time, the sensitivity fields 
already saturate. We can estimate by extrapolation that $\norm{v_{\rho}}, \norm{v_{\rho u}} > 10^5$ if computed using the 
tangent equation, at $t = 115 \, c / a_{\infty}$.

One expects the bias in the ES estimator to be non-negligible for trajectory lengths shorter 
than $t = 115 \, c/a_{\infty}$. That is, the sensitivities must be computed 
at least until the time the information
about the perturbation propagates to the airfoil,
in order for the average sensitivity of the lift to converge to its true value. In other words, convergence 
of the bias requires the entropy wave to reach the airfoil and thus 
$\tau$ must be larger than $\tau^* := 115 \, c / a_{\infty}$. 
In order to predict the cost of the ES method, let us 
make the optimistic assumptions that at $\tau^*$, the bias term is close to zero and that the CLT holds for the variance. A random tangent field is at least $\sim 10^5$ in magnitude at $\tau^*$ implying that 
the variance at $\tau^*$ would be ${\cal O}(10^{10})$. Therefore, under the CLT assumption, which we have shown 
to be too optimistic in our previous examples in sections \ref{sec:lorenz63} and \ref{sec:lorenz96},  
one would require on the order of $10^{10}$ samples for an ${\cal O}(1)$ mean squared error. Solving the primal and the perturbation equations  
on the order of 10 billion times for trajectories of lengths $\tau^*$ would be 
computationally infeasible. This example illustrates that the much shorter timescale of divergence of the adjoint/tangent fields,
compared to that of the convergence of the bias, leads to computational intractability of the ES method. 
\subsection{Turbulent flow over a turbine vane}
\label{sec:vane}
\begin{figure}
\begin{subfigure}{0.5\textwidth}
		\includegraphics[height=6cm]{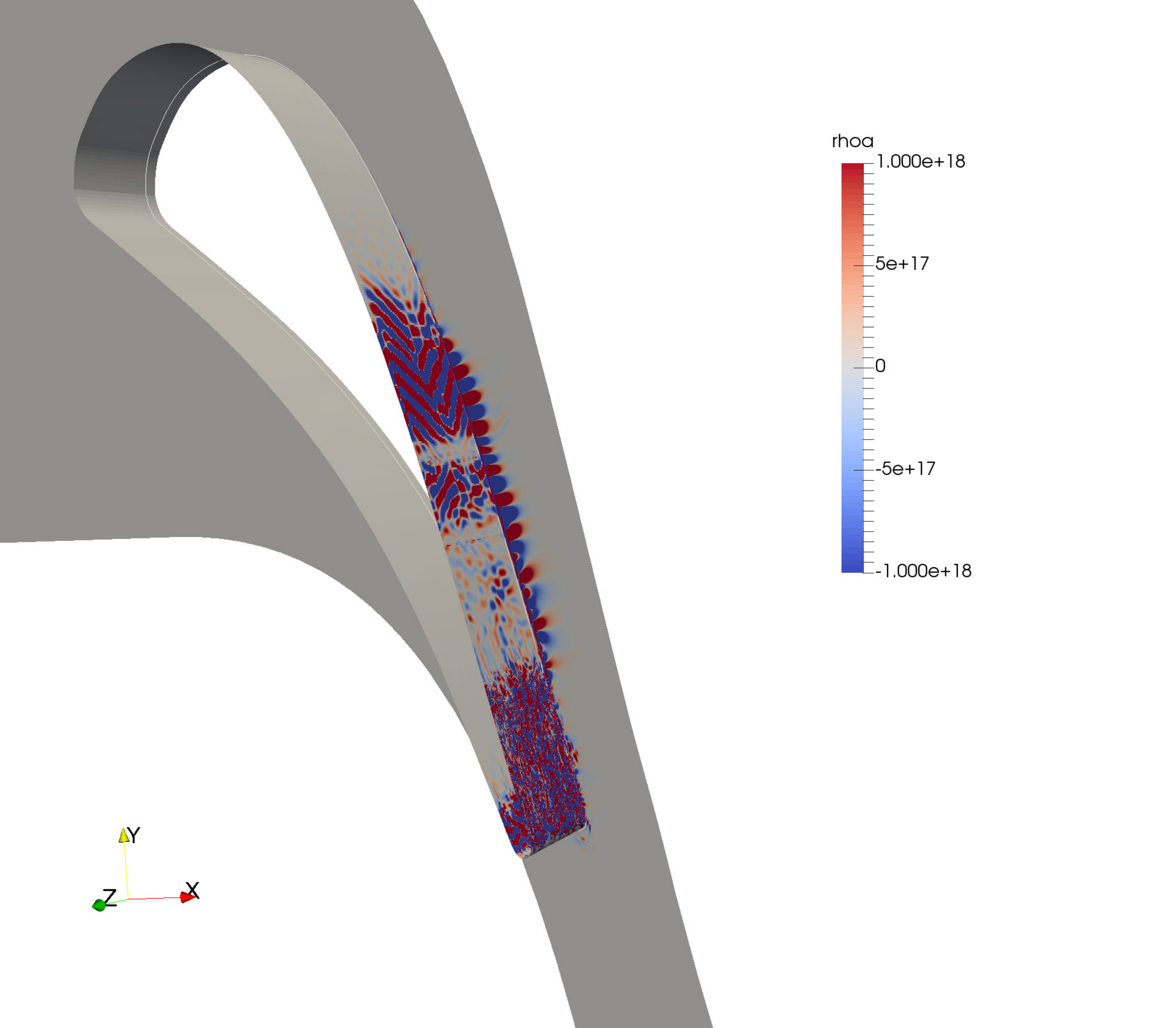}
\caption{Adjoint field corresponding to the density, labelled
		rhoa in the colormap, at $t^* =0.35$. }
\label{fig:vaneAdjointColormap}
\end{subfigure}
\begin{subfigure}{0.45\textwidth}
		\includegraphics[height=6cm,width=\textwidth]{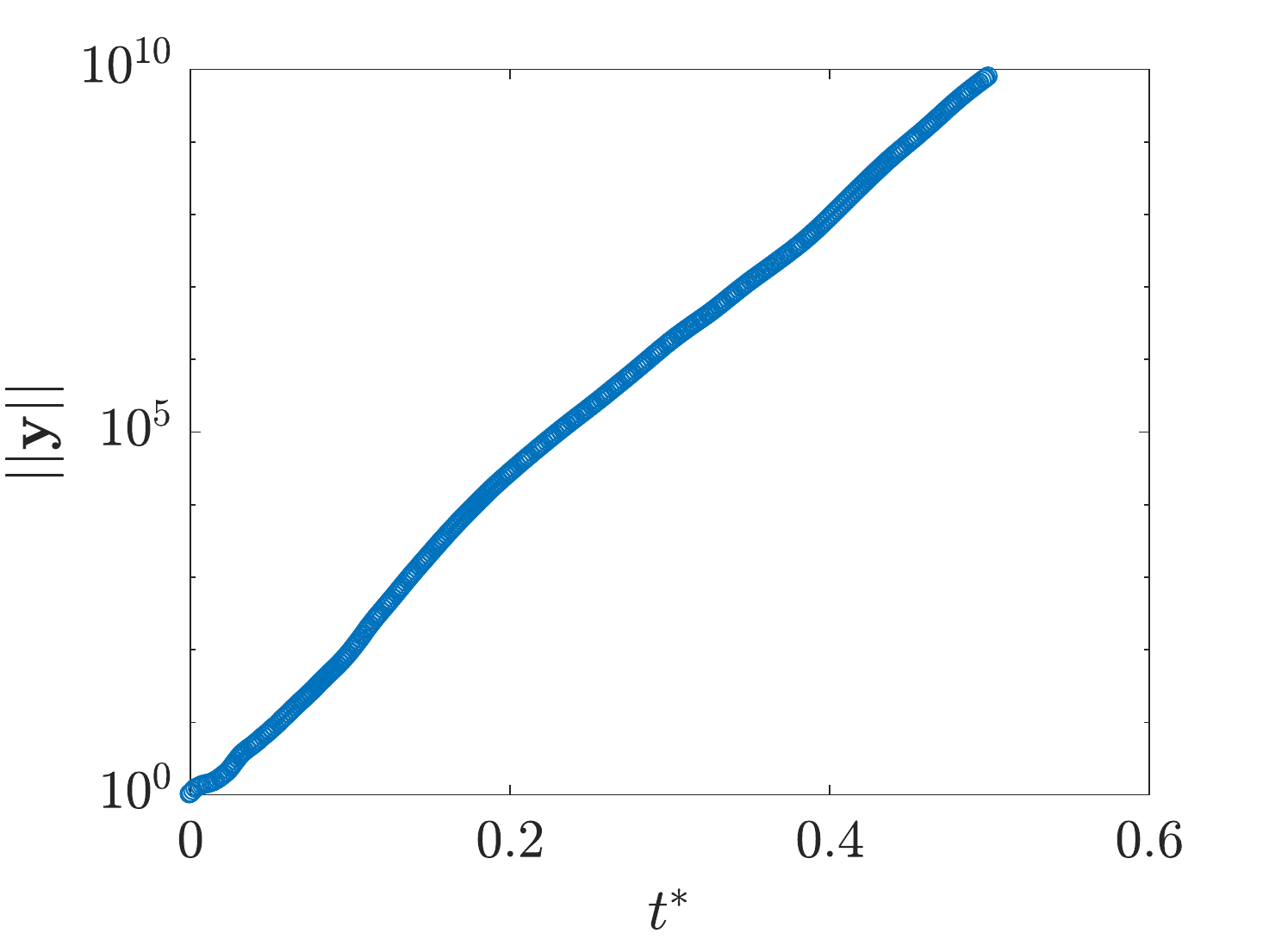}
\caption{The $L^2$ norm of the adjoint vector field, $y$, 
		for the Navier-Stokes system in section \ref{sec:vane} 
		as a function of 
time. }
\label{fig:vaneAdjoint}
\end{subfigure}
\end{figure}

%\begin{multicols}{2}

As a final example, we consider an implicit LES
of turbulent flow around a highly loaded turbine nozzle guide 
	vane performed by Talnikar \emph{et al.} \cite{chai}. Our simulations 
approximate the aero-thermal
experimental investigations of turbine guide vanes in a linear cascade arrangement at the von Karman Institute for Fluid Dynamics 
\cite{vki}. The linear cascade is approximated in the simulation domain 
using periodic boundary conditions in the transverse and
spanwise directions. The spanwise extent, is restricted to 
$0.15$ times the chord length,
which has been numerically verified to be sufficient to capture the turbulent
flow physics. The Reynolds
number of the flow is $10^6$. 
The isentropic Mach number computed with 
respect to the static pressure at the outlet is 0.9.
Isothermal boundary condition is used on the surface of the vane. The fluid achieves transonic speeds as it flows over the surface of
the vane, transitioning to turbulence on the suction side. 

The primal problem is a compressible Navier-Stokes system as in \ref{sec:airfoil},
solved here using
a second order finite volume scheme \cite{chai}. A strong stability-preserving 
third order Runge-Kutta scheme is used for time integration
	and a weighted essentially non-oscillatory scheme \cite{weno}
	is used for shock capturing. The mesh is generated
by uniformly extruding a two-dimensional hybrid structured/unstructured 
mesh in the spanwise direction.

The long-time averaged objective of
interest is the mass-flow averaged pressure loss coefficient 
on a plane 0.25 chord lengths downstream of the trailing edge
of the vane. The reduction in stagnation pressure loss is due to mixing in the turbulent wake and the formation of a
boundary layer on the surface of the vane.
Sensitivities (computed using the adjoint method) are 
calculated with respect to 
Gaussian-shaped source term
perturbations to the system dynamics centered $0.3$ chord lengths upstream 
from the leading edge of the vane in the axial direction. 
The time variable, $t^*$, is nondimensionalized 
with respect to the flow-through
time, which is the duration of the time taken for fluid flow from the inlet to outlet. In  \ref{fig:vaneAdjointColormap}, 
we show the adjoint field corresponding to the density, denoted by \verb+rhoa+, at $t^* = 0.35$.  \ref{fig:vaneAdjoint} shows the $L^2$ norm of the 
adjoint field, denoted by $y$, as a function of $t^*$. 
From the growth of the
adjoint vectors in  \ref{fig:vaneAdjoint}, we can estimate the leading 
Lyapunov exponent for the flow to be $\sim 46$ in nondimensional  
units. The time taken for the entropy wave of the adjoint solution 
to reach the source of the perturbation (the propagation 
direction of the adjoint 
solution is reversed with respect to the primal flow) is 
$\gtrsim 0.5$ time units. From  \ref{fig:vaneAdjoint}, we 
can see that the adjoint solution has
diverged by $10$ orders of magnitude by this time.
	
The reason
for the rapid divergence is the high angle of attack of the
flow onto the vane surface which causes the Mach wave of
the adjoint solution to propagate
with a maximum speed of Mach 1.9 upstream in the axial direction. 
The entropy wave, in comparison, has a lower maximum speed of Mach 0.9, which
is equal to the flow speed. The adjoint diverges quickly once the Mach wave reaches the turbulent 
boundary layer region close to the trailing
edge of the vane on the suction side, as shown in  \ref{fig:vaneAdjointColormap}.
But the bias in the adjoint computed
over an ensemble of trajectories can be expected to
decline only for trajectory lengths $\gtrsim 0.5$ time
units, which is approximately the time required for the entropy
wave to reach the vane leading edge. Similar to the flow over the 
airfoil presented in section \ref{sec:airfoil}, 
the lower speed of propagation of the entropy wave from the 
inlet makes the ES method infeasible in this problem.

Equipped with the estimate of the Lyapunov exponent 
and the timescale of convergence of the bias, we 
predict the cost of the ES method in this case. 
As in section \ref{sec:airfoil}, let us assume the 
best case scenario for both the bias and the variance 
terms. Suppose the bias has negligible magnitude 
at $0.5$ time units and that the CLT assumption is valid.
Since the true variance would be on the order of 
$10^{20}$ at $t^* = 0.5$, we would require ${\cal O}(10^{20})$ 
samples in order to reduce the variance of $\theta_{0.5,N}^{\rm A}$
and consequently, the mean squared error to ${\cal O}(1)$.
To roughly estimate the computational power required, 
we need about $10$ teraFLOPS for $12$ hours for a single adjoint simulation up to 
$t^* = 0.5$; this means approximately $10^{15}$ exaFLOPS for $12$ hours would be needed 
for convergence. This is beyond the computational capabilities 
achievable in the near future. To conclude our
discussion, even though the source of perturbations is close to the leading
edge of the vane, turbulence at a high Reynolds number 
makes the timescale of the growth of perturbations much shorter
than the time required for the propagation of the information 
about the perturbation through entropy waves. As a result, 
since the number of samples required for convergence increases
exponentially at best with time, the ES method becomes computationally intractable.
%https://arxiv.org/pdf/1205.5516.pdf
%the tangent field has diverged by 5 orders of magnitude by then.

Before closing, we remark on the differences between
this example and the previous one in section \ref{sec:airfoil}. 
Although the timescale discrepancy argument supplied to rule 
out the applicability is the same in the two examples, the 
appearance of such a discrepancy is attributed to rather different
reasons. In the airfoil flow, one could have argued that 
the particular choice of the parameter perturbation prolonged 
the timescale for the decay of the bias, leading to the divergence
of the tangent/adjoint within that timescale. That situation 
is typical in an external flow, where parameters we can 
influence are at the farfield, spatially separated 
from quantities of interest. The flow over the turbine vane
in this section, on the
other hand, is an internal flow where the 
flow physics (the slowness of the mean flow
compared to the Mach wave) was responsible 
for the inapplicability of the ES method, rather 
than the limited choice of 
controllable parameters.
\section{Discussion and comments}
\label{sec:discussion}
From the optimistic analysis presented in section \ref{sec:analysis}
and the numerical examples considered in the previous section,
one can conclude that the ES method cannot be a universal
chaotic sensitivity analysis method, owing to its computational cost. 
The main basis for this conclusion is that the ES estimator's
bias often decays at a longer 
timescale than the inverse of the 
largest Lyapunov exponent, which dictates its variance. The analyses and the examples
do not preclude however, cases where this timescale
discrepancy is not large enough to make the ES method
computationally impractical. This leads to the question
of whether one can choose specific objective functions
and parameters, that are dissimilar to the examples
in section \ref{sec:results}, so as to yield 
better scenarios for the application of the ES method. 
Here we provide a closer examination of this
difficult and problem-specific question; we draw upon
connections between signature timescales,
that could guide us in the determination of 
the edge cases for the method's applicability.

As noted in section \ref{sec:analysis}, in uniformly hyperbolic 
systems that we consider here, the decay of 
correlations is exponential. Using the 
Collet-Eckmann conjecture, we argued that the 
rate of the bias decay ($=\gamma_1$) is similar to 
the rate of correlation decay. This argument
is also exemplified, in addition to our heuristic
explanation considering only stable perturbations
in section \ref{sec:optimisticEstimates},
through alternative derivations of Ruelle's response formula
that have appeared in a few
works in the dynamical systems literature (\cite{butterley,
baladi}; see also \cite{nisha} or proposition 8.1 of \cite{gouezel}
for a computable modification
of Ruelle's formula in discrete-time systems). The thrust of these alternative formulations
is that one can replace the computation of Ruelle's formula 
which, in its original form (Eq. \ref{eqn:ruelleLinearResponse}) 
is a time integral with an exponentially increasing variance, 
with a computation that has bounded variance at all times.
These formulae have the convergence of a typical Monte Carlo 
estimator (at the rate of $1/\sqrt{N}$), independent 
of the objective function,
and the bias in these formulae decay as the time correlations
in the system. This implies that, 
supposing $\theta_{\tau,N}$ behaved like a Monte Carlo
estimator for a specific objective function (or only slightly
worse, as in the Lorenz'63 case), the bias decay rate is still bounded 
above by the correlation decay rate.

It seems unintuitive that there exists a connection between the 
two timescales. After all, the two exponential rates, of
the decay of correlations and that of the Lyapunov exponents, 
represent disconnected chaotic
phenomena, namely, the rate of loss of information carried
by state functions, and local expansion of infinitesimal
perturbations, respectively.  Nevertheless, several works have 
verified this connection in low-dimensional uniformly hyperbolic dynamics (see 
\cite{slipantschuk,mendes} for recent work). In fact, numerical 
studies in disparate turbulent flows such as drift wave turbulence in plasma 
\cite{pedersen} and in fluid convection \cite{sirovich}, also 
confirm that the correlations decay slower than $(1/\lambda_1)$. Given the validity 
of the chaotic hypothesis in many turbulent flows, there is reason to expect this 
connection between the two timescales, that can be rigorously shown to hold
only in mathematical toy models, to also manifest itself in general turbulent flows.

At the same time, the rate of correlation decay depends on the observables considered;
this is why one cannot exclude the existence of objective function-parameter pairs 
that lead to fast convergence of $\theta_{\tau,N}$, even though the decay of 
correlations for the system as a whole is at a slower rate than $\lambda_1$ (i.e., 
even if the above conjecture holds). The decay rate of correlations for 
a chaotic system is generally determined 
as the average decay between two functions belonging to a broad function
class (such as the set of continuously differentiable functions of 
the state). There exist observables correlations between which decay 
at a faster rate than that given by this signature rate of the 
system. The correlation time at which the bias converges
is rather difficult to predict a priori. In 
accordance with the approach of this work which isolates extreme 
scenarios, we can take the autocorrelation times of the 
worst-case observables (or the ones with notably slow decay 
rates) as a crude approximation of the upper bound
on the bias decay time. In 
the event that this upper bound is smaller than the inverse 
of the largest Lyapunov exponent, we can expect the ES method 
to behave at least as well as a Monte Carlo estimator. 

In aeroacoustics, an important application area where RANS modeling
is inadequate and turbulence-resolving LES/DNS is resorted to, 
farfield noise (away from a nozzle) 
caused by a turbulent jet is often used as an objective function.
In such a scenario, a wealth of spectral studies (for brevity,
we refer only to a few recent works on modelling turbulent jets 
\cite{schmidt, bodony, freund}) have found that the energy-carrying
spatial Fourier modes (of lower wave number) appear as large-scale
structures with high spatial coherence. This also translates into
high coherence in time or equivalently, a slow decay of time correlations 
of streamwise observables -- this being corroborated by more 
sophisticated mechanisms of turbulent structure 
transit than the oft-used Taylor's hypothesis of frozen turbulence
(among the vast body of work on space-time correlations in turbulent
flow, we point the reader to some comprehensive reviews in 
\cite{hereview, robinson, holmes} that detail this point). 
So, in this case, we can take the integral timescales reported for farfield 
streamwise measurements as the worst-case estimate for the 
bias decay time. Similarly, in turbulent 
boundary layers, integral timescales associated with 
the intermittent or outer regions of the flow can be 
taken as an upper bound since very near-wall regions exhibit 
faster decay of correlations. In the case that
the slowest rate of correlation decay estimated 
as the inverse of this upper bound, 
is still larger than the largest Lyapunov exponent, the ES method 
could be practical.

Given an objective function of interest
and a fixed set of parameters, one might in practice, benefit from 
knowing the decay of correlations from thorougly documented
computational DNS/LES datasets or experimental observations for the 
appropriate type of flow (for example, see \cite{freund, schmidt, bodony} for the 
case of jets and \cite{moin,blackwelder,baars} for the case of turbulent boundary layers). 
In this regard, the availability of space correlation
data, which are more naturally obtained and 
commonly reported in computational studies,
should also prove to be sufficient for estimation
purposes since they can be transformed approximately to time-correlation 
data. If an estimate of the largest 
Lyapunov exponent is also available, one needs to simply compare 
the timescales and use the results of section \ref{sec:analysis}
to predetermine if the ES method would be applicable.

Ultimately, these 
heuristic predictions based on prior knowledge of 
timescales can only help us identify the corner 
cases of applicability (or the lack thereof) of the ES method.
We contend that it is more likely to encounter undecidable cases
and close this discussion with such an example that appears 
in Rayleigh-B\'enard convection. 
Sirovich \emph{et al.} \cite{sirovich} 
have shown that the correlation time defined
as the time it takes for the streamwise velocity autocorrelations
to drop to its first minimum, is on the same order 
of magnitude as the largest Lyapnov exponent. Interestingly,
our numerical results for the Lorenz'63 system (section \ref{sec:lorenz63}), 
which is a mathematical model for Rayleigh-B\'enard convection, 
is also in line with this finding -- the rate of decay of the bias 
and the largest (and only)
positive Lyapunov exponent are both ${\cal O}(1)$. In such a scenario, the practitioner
may choose to rely on alternative methods \cite{angxiu,nisha,lucarini,
patrick} in order to avoid 
incurring the cost of solving an enormous number of tangent or 
adjoint equations without being to establish convergence. 
\section{Conclusion}
\label{sec:conclusion}
To compute sensitivities with respect to design parameters,
of statistically stationary quantities in chaotic systems,
ensemble methods appear to be an appealing solution; they 
are both conceptually simpler and easier to implement than 
fluctuation-dissipation-based \cite{lucarini} and shadowing-based \cite{angxiu_nilsas}
methods, all of which are, moreover, still under active development. 
However, the present work has shown that Eyink \emph{et al.}'s \cite{eyink}
results revealing poor convergence of ensemble computations in the Lorenz'63
system, is more widely representative of convergence trends in general chaotic 
systems. The present work deals with estimating
a theoretical upper bound on the rate of 
convergence of an ensemble sensitivity, agnostic to the 
objective function-parameter pair.

For this, the most optimistic assumptions are made 
on the bias and variance associated with the ensemble
sensitivity (ES) estimator, under the mathematical 
simplification of uniform hyperbolicity. We show that, with 
the integration time, in the best case, the bias decays 
exponentially at a problem-dependent rate and the variance 
increases exponentially at the rate of twice the largest Lyapunov exponent. 
Assuming these optimistic bounds, the computational cost of 
the ES method, in theory, still scales exponentially with the mean squared error.
The ratio of the bias decay rate to the leading Lyapunov exponent of the system, is 
the single most influential parameter that determines the 
feasibility of the method.

Our numerical results for the Lorenz'63 system show that 
the optimistic model proposed for the least mean squared error
is only locally applicable. The upper bound on the rate of convergence 
is 0.5, concurring with Eyink \emph{et al.}'s \cite{eyink} results. The 40-variable 
Lorenz'96 system serves as an example of a low-dimensional
attractor for which the asymptotic
convergence is remarkably slow. Although the rate of convergence
is poor for this system, the mean squared error magnitudes
were low at a reasonable computational cost, for the chosen objective
function. This suggests that one may encounter, in practice, 
objective functions for which the ensemble sensitivities are within
a specified accuracy, at an affordable computational cost. 
In the numerical simulations of chaotic fluid flow we consider, 
we obtain optimistic estimates
on the rate of convergence which hold true for a general objective function.
Our results indicate that the flow physics imposes an 
upper bound on the rate of convergence. Altogether, 
the present numerical evidence suggests the following: 
even under the optimistic assumption of exponential decay of the bias, 
the cost of exponential sampling of an expensive primal 
problem can make the ES method infeasible in practical applications, 
for a general objective function-parameter pair. However, there may 
exist objective function and parameter choices that lead to a 
smaller timescale discrepancy between the bias decay and 
the variance increase -- this could lead to a faster convergence 
than the estimate of the upper bound.
\section*{Funding Sources}
 This work was supported by AFOSR Award
FA9550-15-1-0072 under Dr. Fariba Fahroo and Dr. Jean-luc Cambier.

\section*{Acknowledgments}
The authors would like to thank the reviewers for their insightful comments and Angxiu Ni for helpful discussions. 
This research used resources of the Argonne Leadership 
Computing Facility, which is a DOE Office of Science User Facility supported under
Contract DE-AC02-06CH11357.
Pablo Fernandez would like to acknowledge financial support from the MIT Zakhartchenko Fellowship.
\bibliography{sample}

\begin{thebibliography}{60}
\newcommand{\enquote}[1]{``#1''}
\providecommand{\natexlab}[1]{#1}
\providecommand{\url}[1]{\texttt{#1}}
\providecommand{\urlprefix}{URL }
\expandafter\ifx\csname urlstyle\endcsname\relax
  \providecommand{\doi}[1]{doi:\discretionary{}{}{}#1}\else
  \providecommand{\doi}{doi:\discretionary{}{}{}\begingroup
  \urlstyle{rm}\Url}\fi

\bibitem[{Palacios et~al.(2012)Palacios, Duraisamy, Alonso, and
  Zuazua}]{palacios}
Palacios, F., Duraisamy, K., Alonso, J.~J., and Zuazua, E., \enquote{Robust
  grid adaptation for efficient uncertainty quantification,} \emph{AIAA
  journal}, Vol.~50, No.~7, 2012, pp. 1538--1546.
\newblock \doi{10.2514/1.J051379}.

\bibitem[{Wang et~al.(2012)Wang, Duraisamy, Alonso, and
  Iaccarino}]{qiqi-unstart}
Wang, Q., Duraisamy, K., Alonso, J.~J., and Iaccarino, G., \enquote{Risk
  assessment of scramjet unstart using adjoint-based sampling methods,}
  \emph{AIAA journal}, Vol.~50, No.~3, 2012, pp. 581--592.
\newblock \doi{10.2514/1.J051264}.

\bibitem[{Fidkowski and Darmofal(2011)}]{fidkowski}
Fidkowski, K.~J., and Darmofal, D.~L., \enquote{Review of output-based error
  estimation and mesh adaptation in computational fluid dynamics,} \emph{AIAA
  journal}, Vol.~49, No.~4, 2011, pp. 673--694.
\newblock \doi{10.2514/1.J050073}.

\bibitem[{Rizzetta and Visbal(2003)}]{rizzetta}
Rizzetta, D.~P., and Visbal, M.~R., \enquote{Large-eddy simulation of
  supersonic cavity flowfields including flow control,} \emph{AIAA journal},
  Vol.~41, No.~8, 2003, pp. 1452--1462.
\newblock \doi{10.2514/2.2128}.

\bibitem[{Bodony and Lele(2008)}]{bodony}
Bodony, D.~J., and Lele, S.~K., \enquote{Current status of jet noise
  predictions using large-eddy simulation,} \emph{AIAA journal}, Vol.~46,
  No.~2, 2008, pp. 364--380.
\newblock \doi{10.2514/1.24475}.

\bibitem[{Engblom et~al.(2004)Engblom, Khavaran, and Bridges}]{engblom}
Engblom, W., Khavaran, A., and Bridges, J., \enquote{Numerical prediction of
  chevron nozzle noise reduction using WIND-MGBK methodology,} \emph{10th
  AIAA/CEAS Aeroacoustics Conference}, 2004, p. 2979.
\newblock \doi{10.2514/6.2004-2979}.

\bibitem[{Tucker(2004)}]{paul}
Tucker, P.~G., \enquote{Novel MILES computations for jet flows and noise,}
  \emph{International Journal of Heat and Fluid Flow}, Vol.~25, No.~4, 2004,
  pp. 625--635.
\newblock \doi{10.1016/j.ijheatfluidflow.2003.11.021}.

\bibitem[{Peter and Dwight(2010)}]{peter}
Peter, J.~E., and Dwight, R.~P., \enquote{Numerical sensitivity analysis for
  aerodynamic optimization: A survey of approaches,} \emph{Computers \&
  Fluids}, Vol.~39, 2010, pp. 373--391.
\newblock \doi{10.1016/j.compfluid.2009.09.013}.

\bibitem[{Giles and Pierce(2000)}]{giles1}
Giles, M.~B., and Pierce, N.~A., \enquote{An introduction to the adjoint
  approach to design,} \emph{Flow, turbulence and combustion}, Vol.~65, 2000,
  pp. 393--415.
\newblock \doi{10.1023/A:1011430410075}.

\bibitem[{Giles et~al.(2003)Giles, Duta, M-uacute, ller, and Pierce}]{giles2}
Giles, M.~B., Duta, M.~C., M-uacute, J.-D., ller, and Pierce, N.~A.,
  \enquote{Algorithm developments for discrete adjoint methods,} \emph{AIAA
  journal}, Vol.~41, No.~2, 2003, pp. 198--205.
\newblock \doi{10.2514/2.1961}.

\bibitem[{RA~Martins et~al.(2004)RA~Martins, Alonso, and Reuther}]{alonso2}
RA~Martins, J.~R., Alonso, J.~J., and Reuther, J.~J., \enquote{High-fidelity
  aerostructural design optimization of a supersonic business jet,}
  \emph{Journal of Aircraft}, Vol.~41, No.~3, 2004, pp. 523--530.
\newblock \doi{10.2514/1.11478}.

\bibitem[{Nielsen and Anderson(1999)}]{nielsen}
Nielsen, E.~J., and Anderson, W.~K., \enquote{Aerodynamic design optimization
  on unstructured meshes using the Navier-Stokes equations,} \emph{AIAA
  journal}, Vol.~37, No.~11, 1999, pp. 1411--1419.
\newblock \doi{10.2514/2.640}.

\bibitem[{Ni and Wang(2017)}]{angxiu}
Ni, A., and Wang, Q., \enquote{Sensitivity analysis on chaotic dynamical
  systems by Non-Intrusive Least Squares Shadowing (NILSS),} \emph{Journal of
  Computational Physics}, Vol. 347, 2017, pp. 56--77.
\newblock \doi{10.1016/j.jcp.2017.06.033}.

\bibitem[{Lea et~al.(2000)Lea, Allen, and Haine}]{lea}
Lea, D.~J., Allen, M.~R., and Haine, T.~W., \enquote{Sensitivity analysis of
  the climate of a chaotic system,} \emph{Tellus A: Dynamic Meteorology and
  Oceanography}, Vol.~52, 2000, pp. 523--532.
\newblock \doi{10.1034/j.1600-0870.2000.01137.x}.

\bibitem[{Capecelatro et~al.(2018)Capecelatro, Bodony, and
  Freund}]{capecelatro}
Capecelatro, J., Bodony, D.~J., and Freund, J.~B., \enquote{Adjoint-based
  sensitivity and ignition threshold mapping in a turbulent mixing layer,}
  \emph{Combustion Theory and Modelling}, 2018, pp. 1--33.
\newblock \doi{10.2514/6.2017-0846}.

\bibitem[{Moigne and Qin(2004)}]{alonso1}
Moigne, A.~L., and Qin, N., \enquote{Variable-fidelity aerodynamic optimization
  for turbulent flows using a discrete adjoint formulation,} \emph{AIAA
  journal}, Vol.~42, No.~7, 2004, pp. 1281--1292.
\newblock \doi{10.2514/1.2109}.

\bibitem[{Ruelle(1997)}]{ruelle}
Ruelle, D., \enquote{Differentiation of SRB states,} \emph{Communications in
  Mathematical Physics}, Vol. 187, 1997, pp. 227--241.
\newblock \doi{10.1007/s002200050134}.

\bibitem[{Eyink et~al.(2004)Eyink, Haine, and Lea}]{eyink}
Eyink, G., Haine, T., and Lea, D., \enquote{Ruelle's linear response formula,
  ensemble adjoint schemes and L{\'e}vy flights,} \emph{Nonlinearity}, Vol.~17,
  2004, p. 1867.
\newblock \doi{10.1088/0951-7715/17/5/016}.

\bibitem[{Katok and Hasselblatt(1997)}]{katok}
Katok, A., and Hasselblatt, B., \emph{Introduction to the modern theory of
  dynamical systems}, Vol.~54, Cambridge university press, 1997.
\newblock \doi{10.1017/CBO9780511809187}.

\bibitem[{Ni(2018)}]{angxiu_nilsas}
Ni, A., \enquote{Sensitivity analysis on chaotic dynamical systems by
  Non-Intrusive Least Squares Adjoint Shadowing (NILSAS),} \emph{arXiv preprint
  arXiv:1801.08674}, 2018.

\bibitem[{Ragone et~al.(2016)Ragone, Lucarini, and Lunkeit}]{lucarini}
Ragone, F., Lucarini, V., and Lunkeit, F., \enquote{A new framework for climate
  sensitivity and prediction: a modelling perspective,} \emph{Climate
  Dynamics}, Vol.~46, 2016, pp. 1459--1471.
\newblock \doi{10.1007/s00382-015-2657-3}.

\bibitem[{Blonigan(2017)}]{patrick}
Blonigan, P.~J., \enquote{Adjoint sensitivity analysis of chaotic dynamical
  systems with non-intrusive least squares shadowing,} \emph{Journal of
  Computational Physics}, Vol. 348, 2017, pp. 803--826.
\newblock \doi{10.1016/j.jcp.2017.08.002}.

\bibitem[{Blonigan et~al.(2017{\natexlab{a}})Blonigan, Fernandez, Murman, Wang,
  Rigas, and Magri}]{patrickPablo}
Blonigan, P.~J., Fernandez, P., Murman, S.~M., Wang, Q., Rigas, G., and Magri,
  L., \enquote{Toward a chaotic adjoint for LES,} \emph{arXiv preprint
  arXiv:1702.06809}, 2017{\natexlab{a}}.

\bibitem[{Blonigan et~al.(2017{\natexlab{b}})Blonigan, Wang, Nielsen, and
  Diskin}]{patrick-aiaa}
Blonigan, P.~J., Wang, Q., Nielsen, E.~J., and Diskin, B.,
  \enquote{Least-Squares Shadowing Sensitivity Analysis of Chaotic Flow Around
  a Two-Dimensional Airfoil,} \emph{AIAA Journal}, 2017{\natexlab{b}}, pp.
  658--672.
\newblock \doi{10.2514/1.J055389}.

\bibitem[{Gallavotti and Cohen(1995)}]{gallavotti}
Gallavotti, G., and Cohen, E., \enquote{Dynamical ensembles in stationary
  states,} \emph{Journal of Statistical Physics}, Vol.~80, 1995, pp. 931--970.
\newblock \doi{10.1007/BF02179860}.

\bibitem[{Ruelle(2003)}]{ruelle1}
Ruelle, D., \enquote{Differentiation of SRB states: correction and
  complements,} \emph{Communications in mathematical physics}, Vol. 234, 2003,
  pp. 185--190.
\newblock \doi{10.1007/s00220-002-0779-z}.

\bibitem[{Ruelle(2008)}]{ruelle_hyperbolicflows}
Ruelle, D., \enquote{Differentiation of SRB states for hyperbolic flows,}
  \emph{Ergodic Theory and Dynamical Systems}, Vol.~28, 2008, pp. 613--631.
\newblock \doi{10.1017/S0143385707000260}.

\bibitem[{Collet and Eckmann(2004)}]{collet}
Collet, P., and Eckmann, J.-P., \enquote{Liapunov multipliers and decay of
  correlations in dynamical systems,} \emph{Journal of statistical physics},
  Vol. 115, No. 1-2, 2004, pp. 217--254.
\newblock \doi{10.1023/B:JOSS.0000019817.71073.61}.

\bibitem[{Baladi et~al.(1989)Baladi, Eckmann, and Ruelle}]{baladi1}
Baladi, V., Eckmann, J.-P., and Ruelle, D., \enquote{Resonances for
  intermittent systems,} \emph{Nonlinearity}, Vol.~2, 1989, p. 119.
\newblock \doi{10.1088/0951-7715/2/1/007}.

\bibitem[{Baladi(2000)}]{baladi2}
Baladi, V., \emph{Positive transfer operators and decay of correlations},
  Vol.~16, World scientific, 2000.
\newblock \doi{10.1142/3657}.

\bibitem[{{Lorenz}(1963)}]{lorenz63}
{Lorenz}, E.~N., \enquote{Deterministic Nonperiodic Flow,} \emph{Journal of
  Atmospheric Sciences}, 1963.
\newblock \doi{10.1175/1520-0469(1963)020<0130:DNF>2.0.CO;2}.

\bibitem[{Holland and Melbourne(2007)}]{lorenz63_CLT}
Holland, M., and Melbourne, I., \enquote{Central limit theorems and invariance
  principles for Lorenz attractors,} \emph{Journal of the London Mathematical
  Society}, Vol.~76, 2007, pp. 345--364.
\newblock \doi{10.1112/jlms/jdm060}.

\bibitem[{Lorenz(1996)}]{lorenz96}
Lorenz, E.~N., \enquote{Predictability: A problem partly solved,} \emph{Proc.
  Seminar on predictability}, Vol.~1, 1996.
\newblock \doi{10.1017/CBO9780511617652.004}.

\bibitem[{Karimi and Paul(2010)}]{karimi}
Karimi, A., and Paul, M.~R., \enquote{Extensive chaos in the Lorenz-96 model,}
  \emph{Chaos: An Interdisciplinary Journal of Nonlinear Science}, Vol.~20,
  2010, p. 043105.
\newblock \doi{10.1063/1.3496397}.

\bibitem[{Venturi et~al.(2016)Venturi, Cho, and Karniadakis}]{mori}
Venturi, D., Cho, H., and Karniadakis, G.~E., \enquote{Mori-Zwanzig Approach to
  Uncertainty Quantification,} \emph{Handbook of Uncertainty Quantification.
  Springer}, 2016.
\newblock \doi{10.1007/978-3-319-11259-6_28-2}.

\bibitem[{Fernandez and Wang(2017)}]{pablo1}
Fernandez, P., and Wang, Q., \enquote{Lyapunov spectrum of the separated flow
  around the NACA 0012 airfoil and its dependence on numerical discretization,}
  \emph{Journal of Computational Physics}, Vol. 350, 2017, pp. 453--469.
\newblock \doi{10.1016/j.jcp.2017.08.056}.

\bibitem[{Pulliam and Vastano(1993)}]{pulliam}
Pulliam, T.~H., and Vastano, J.~A., \enquote{{Transition to Chaos in an Open
  Unforced 2D Flow},} \emph{Journal of Computational Physics}, Vol. 105, 1993,
  pp. 133--149.
\newblock \doi{10.1006/jcph.1993.1059}.

\bibitem[{Fernandez(2018)}]{pablo2}
Fernandez, P., \enquote{Entropy-stable hybridized discontinuous Galerkin
  methods for large-eddy simulation of transitional and turbulent flows,} Ph.D.
  thesis, Massachusetts Institute of Technology, 2018.

\bibitem[{Fernandez et~al.(2017{\natexlab{a}})Fernandez, Nguyen, and
  Peraire}]{pablo3}
Fernandez, P., Nguyen, N., and Peraire, J., \enquote{The hybridized
  Discontinuous Galerkin method for Implicit Large-Eddy Simulation of
  transitional turbulent flows,} \emph{Journal of Computational Physics}, Vol.
  336, 2017{\natexlab{a}}, pp. 308--329.
\newblock \doi{10.1016/j.jcp.2017.02.015}.

\bibitem[{Fernandez et~al.(2017{\natexlab{b}})Fernandez, Nguyen, and
  Peraire}]{Fernandez:2017:AIAA:SGS}
Fernandez, P., Nguyen, N., and Peraire, J., \enquote{Subgrid-scale modeling and
  implicit numerical dissipation in DG-based Large-Eddy Simulation,} \emph{23rd
  AIAA Computational Fluid Dynamics Conference}, 2017{\natexlab{b}}.
\newblock \doi{10.2514/6.2017-3951}.

\bibitem[{Alexander(1977)}]{Alexander:77}
Alexander, R., \enquote{Diagonally implicit Runge--Kutta methods for stiff
  ODE’s,} \emph{SIAM Journal on Numerical Analysis}, Vol.~14, 1977, pp.
  1006--1021.
\newblock \doi{10.1137/0714068}.

\bibitem[{Talnikar et~al.(2017)Talnikar, Wang, and Laskowski}]{chai}
Talnikar, C., Wang, Q., and Laskowski, G.~M., \enquote{Unsteady adjoint of
  pressure loss for a fundamental transonic turbine vane,} \emph{Journal of
  Turbomachinery}, Vol. 139, 2017, p. 031001.
\newblock \doi{10.1115/1.4034800}.

\bibitem[{Arts and De~Rouvroit(1990)}]{vki}
Arts, T., and De~Rouvroit, M.~L., \enquote{Aero-thermal performance of a two
  dimensional highly loaded transonic turbine nozzle guide vane: A test case
  for inviscid and viscous flow computations,} \emph{ASME 1990 International
  Gas Turbine and Aeroengine Congress and Exposition}, American Society of
  Mechanical Engineers, 1990, pp. V001T01A106--V001T01A106.
\newblock \doi{10.1115/1.2927978}.

\bibitem[{Liu et~al.(1994)Liu, Osher, and Chan}]{weno}
Liu, X.-D., Osher, S., and Chan, T., \enquote{Weighted essentially
  non-oscillatory schemes,} \emph{Journal of computational physics}, Vol. 115,
  1994, pp. 200--212.
\newblock \doi{10.1006/jcph.1994.1187}.

\bibitem[{Butterley and Liverani(2007)}]{butterley}
Butterley, O., and Liverani, C., \enquote{Smooth Anosov flows: correlation
  spectra and stability,} \emph{J. Mod. Dyn}, Vol.~1, No.~2, 2007, pp.
  301--322.
\newblock \doi{10.3934/jmd.2007.1.301}.

\bibitem[{Baladi(2018)}]{baladi}
Baladi, V., \emph{Dynamical zeta functions and dynamical determinants for
  hyperbolic maps}, Springer, 2018.
\newblock \doi{10.1007/978-3-319-77661-3}.

\bibitem[{Chandramoorthy et~al.(2019)Chandramoorthy, Wang, Wang, and
  Tucker}]{nisha}
Chandramoorthy, N., Wang, Z.-N., Wang, Q., and Tucker, P., \enquote{Toward
  computing sensitivities of average quantities in turbulent flows,}
  \emph{arXiv preprint arXiv:1902.11112}, 2019.

\bibitem[{Gou{\"e}zel et~al.(2008)Gou{\"e}zel, Liverani et~al.}]{gouezel}
Gou{\"e}zel, S., Liverani, C., et~al., \enquote{Compact locally maximal
  hyperbolic sets for smooth maps: fine statistical properties,} \emph{Journal
  of Differential Geometry}, Vol.~79, No.~3, 2008, pp. 433--477.
\newblock \doi{10.4310/jdg/1213798184}.

\bibitem[{Slipantschuk et~al.(2013)Slipantschuk, Bandtlow, and
  Just}]{slipantschuk}
Slipantschuk, J., Bandtlow, O.~F., and Just, W., \enquote{On the relation
  between Lyapunov exponents and exponential decay of correlations,}
  \emph{Journal of Physics A: Mathematical and Theoretical}, Vol.~46, No.~7,
  2013, p. 075101.
\newblock \doi{10.1088/1751-8113/46/7/075101}.

\bibitem[{Mendes et~al.(2019)Mendes, da~Silva, and Beims}]{mendes}
Mendes, C., da~Silva, R., and Beims, M., \enquote{Decay of distance
  autocorrelation and Lyapunov exponents,} \emph{arXiv preprint
  arXiv:1903.08202}, 2019.

\bibitem[{Pedersen et~al.(1996)Pedersen, Michelsen, and Rasmussen}]{pedersen}
Pedersen, T.~S., Michelsen, P.~K., and Rasmussen, J.~J., \enquote{Lyapunov
  exponents and particle dispersion in drift wave turbulence,} \emph{Physics of
  Plasmas}, Vol.~3, No.~8, 1996, pp. 2939--2950.
\newblock \doi{10.1063/1.871636}.

\bibitem[{Sirovich and Deane(1991)}]{sirovich}
Sirovich, L., and Deane, A.~E., \enquote{A computational study of
  Rayleigh--B{\'e}nard convection. Part 2. Dimension considerations,}
  \emph{Journal of fluid mechanics}, Vol. 222, 1991, pp. 251--265.
\newblock \doi{10.1017/S002211209100109X}.

\bibitem[{Schmidt et~al.(2018)Schmidt, Towne, Rigas, Colonius, and
  Br{\`e}s}]{schmidt}
Schmidt, O.~T., Towne, A., Rigas, G., Colonius, T., and Br{\`e}s, G.~A.,
  \enquote{Spectral analysis of jet turbulence,} \emph{Journal of Fluid
  Mechanics}, Vol. 855, 2018, pp. 953--982.
\newblock \doi{10.1017/jfm.2018.675}.

\bibitem[{Freund et~al.(2000)Freund, Lele, and Moin}]{freund}
Freund, J., Lele, S., and Moin, P., \enquote{Numerical simulation of a Mach
  1.92 turbulent jet and its sound field,} \emph{AIAA journal}, Vol.~38,
  No.~11, 2000, pp. 2023--2031.
\newblock \doi{10.2514/2.889}.

\bibitem[{He et~al.(2017)He, Jin, and Yang}]{hereview}
He, G., Jin, G., and Yang, Y., \enquote{Space-time correlations and dynamic
  coupling in turbulent flows,} \emph{Annual Review of Fluid Mechanics},
  Vol.~49, 2017, pp. 51--70.
\newblock \doi{10.1146/annurev-fluid-010816-060309}.

\bibitem[{Robinson(1991)}]{robinson}
Robinson, S.~K., \enquote{Coherent motions in the turbulent boundary layer,}
  \emph{Annual Review of Fluid Mechanics}, Vol.~23, No.~1, 1991, pp. 601--639.
\newblock \doi{10.1146/annurev.fl.23.010191.003125}.

\bibitem[{Holmes et~al.(2012)Holmes, Lumley, Berkooz, and Rowley}]{holmes}
Holmes, P., Lumley, J.~L., Berkooz, G., and Rowley, C.~W., \emph{Turbulence,
  coherent structures, dynamical systems and symmetry}, Cambridge university
  press, 2012.
\newblock \doi{10.1017/CBO9780511919701}.

\bibitem[{Moin and Kim(1982)}]{moin}
Moin, P., and Kim, J., \enquote{Numerical investigation of turbulent channel
  flow,} \emph{Journal of fluid mechanics}, Vol. 118, 1982, pp. 341--377.
\newblock \doi{10.1017/S0022112082001116}.

\bibitem[{Blackwelder and Kovasznay(1972)}]{blackwelder}
Blackwelder, R.~F., and Kovasznay, L.~S., \enquote{Time scales and correlations
  in a turbulent boundary layer,} \emph{The Physics of Fluids}, Vol.~15, No.~9,
  1972, pp. 1545--1554.
\newblock \doi{10.1063/1.1694128}.

\bibitem[{Baars et~al.(2017)Baars, Hutchins, and Marusic}]{baars}
Baars, W., Hutchins, N., and Marusic, I., \enquote{Reynolds number trend of
  hierarchies and scale interactions in turbulent boundary layers,}
  \emph{Philosophical Transactions of the Royal Society A: Mathematical,
  Physical and Engineering Sciences}, Vol. 375, No. 2089, 2017, p. 20160077.
\newblock \doi{10.1098/rsta.2016.0077}.

\end{thebibliography}

\end{document}